\documentclass[11pt]{article}
\usepackage{makeidx}
\usepackage[utf8]{inputenc} 
\usepackage{times}
\usepackage{bm}
\usepackage{comment}
\usepackage{amssymb}
\usepackage{amsmath}
\usepackage{enumitem} 
\usepackage{pdflscape}
\usepackage{listings}
\numberwithin{equation}{section}
\usepackage{color}
\usepackage{graphicx}
\usepackage{rotating}
\usepackage{pdflscape}
\usepackage{epstopdf}
\usepackage[round,authoryear,comma]{natbib}
\usepackage{natbib,hyperref}
\usepackage{xurl}
\usepackage{booktabs}
\usepackage{actuarialangle}
\usepackage{actuarialsymbol}
\usepackage{setspace}
\usepackage{physics}
\usepackage{xcolor}
\usepackage{multirow}
\usepackage{caption}
\usepackage{authblk}
\usepackage{csquotes} 
\usepackage{bbm}
\usepackage[lofdepth,lotdepth]{subfig}
\usepackage{tabularx}
\usepackage{booktabs}
\usepackage{float}
\newcolumntype{Y}{>{\centering\arraybackslash}X} 
\newcolumntype{s}{>{\centering\arraybackslash\hsize=.3\hsize}X} 
\hypersetup{
	colorlinks   = true, 
	urlcolor     = blue, 
	linkcolor    = blue, 
	citecolor   = blue 
}

\definecolor{OliveGreen}{rgb}{0,0.6,0}
\definecolor{alizarin}{rgb}{0.82, 0.1, 0.26}

\providecommand{\U}[1]{\protect\rule{.1in}{.1in}}
\parskip=\medskipamount

\defcitealias{centersfordiseasecontrolandpreventioncdc1984}{CDC, 1984}

\begin{document}


\title{Bayesian mortality modelling with pandemics: a vanishing jump approach}

\author[1]{Julius Goes \thanks{julius.goes@uni-bamberg.de}}
\author[2]{Karim Barigou\thanks{karim.barigou@uclouvain.be}}
\author[1]{Anne Leucht\thanks{anne.leucht@uni-bamberg.de}}
\affil[1]{University of Bamberg, Institute of Statistics, Bamberg, Germany}
\affil[2]{LIDAM-ISBA, Université Catholique de Louvain, Voie du Roman Pays 20, Louvain-La-Neuve, 1348, Belgium}

\date{Version: \today }

\maketitle

\begin{abstract}
This paper extends the Lee-Carter model for single- and multi-populations to account for pandemic jump effects of vanishing kind, allowing for a more comprehensive and accurate representation of mortality rates during a pandemic, characterised by a high impact at the beginning and gradually vanishing effects over subsequent periods. While the Lee-Carter model is effective in capturing mortality trends, it may not be able to account for large, unexpected jumps in mortality rates caused by pandemics or wars. Existing models allow either for transient jumps with an effect of one period only or persistent jumps. However, there is no literature on estimating mortality time series with jumps having an effect over a small number of periods as typically observed in pandemics. The Bayesian approach allows to quantify the uncertainty around the parameter estimates. Empirical data from the COVID-19 pandemic shows the superiority of the proposed approach, compared to models with a transitory shock effect. 
\medskip


\end{abstract}

\section{Introduction}
The model of \cite{lee1992modeling} has been widely used in actuarial science and demography to forecast mortality rates based on past observations. This model assumes that mortality rates follow a stochastic trend with a time-dependent mortality factor, adjusted for age-specific effects, using two sets of age-dependent coefficients. While this model has shown to be effective in capturing mortality trends in many countries, it may not be able to adequately account for large, unexpected jumps in mortality rates, such as those caused by pandemics \citep{chen2022modeling,van2022estimating}. This paper aims to fill the gap by proposing a new modelling framework that is suitable in capturing the typical effects of a pandemic on the subsequent age-specific mortality rates. More specifically, we introduce single and multi-population models that integrate serial dependence via vanishing shocks, where the shock’s impact is highest in the beginning and then gradually diminishes in the subsequent years, allowing for a more reasonable description of the mortality rates during pandemics and a superior model fit.

We adopt a Bayesian framework for mortality modelling, a decision driven by the inherent advantages of this approach. Specifically, the Bayesian methodology integrates the estimation and forecasting processes, ensuring more consistent and robust estimates as underscored by \cite{cairns2011bayesian} and \cite{wong2018bayesian}. Furthermore, this approach excels in accommodating various sources of uncertainty in a coherent manner. Within the literature on Bayesian mortality modelling, \cite{czado2005bayesian} pioneered the application of a comprehensive Bayesian approach specific to the Poisson Lee–Carter (LC) model. This methodology was later expanded to a multi-population context by \cite{antonio2015bayesian}. Other notable contributions include \cite{pedroza2006bayesian}, who employed a Bayesian state-space model using Kalman filters to address missing data issues in mortality forecasting. The development and wider availability of Markov chain Monte Carlo (MCMC) techniques have further increased the application of Bayesian methodologies in mortality modelling, as evidenced by the growing body of recent work including \citep[e.g.][]{alexopoulos2019bayesian,li2019bayesian, barigou2023bayesian, wong2023bayesian}.

The COVID-19 pandemic has emphasised the importance of incorporating mortality shocks into the LC model. While jump effects have been introduced in previous studies, they often fail to account for age-specific pandemic effects. \cite{cox2006multivariate} and \cite{chen2009modeling} proposed extensions with permanent and transitory jump effects, respectively, but the jump effect is applied to the time-dependent factor instead of the mortality rates. This means, that the age pattern of a potential shock is identical to that of the general mortality improvement. To address this shortcoming, \cite{liu2015age} extended the Lee-Carter model by including a time- and age-dependent jump effect, which allows, for example, to capture the age-specific effect of COVID-19. However, their method only allows the inclusion of transitory mortality shocks that last one period, i.e. one year. Here, these effects are incorporated using independent and identically distributed (i.i.d.) shock variables. Consequently, during years of pandemic, mortality rates experience an upward shift. However, this model has a limitation in its assumption of a time-independent jump variable. Specifically, it implies that during years with consecutive shocks, the severity of mortality rate adjustments is independent, or for years following a single-period shock, the effect completely vanishes in the subsequent year. Such assumptions are inconsistent with observed pandemic patterns, wherein mortality rates are heavily impacted during initial stages and gradually taper off.  

In the aftermath of the COVID pandemic, a variety of models were proposed to capture the nuances of mortality trends. For example, \cite{van2022estimating} extended the multi-population model originally proposed by \cite{li2005coherent} for this purpose. Their model integrates three layers: the first two focusing on pre-COVID mortality trends and the third specifically capturing the excess mortality attributable to COVID. The framework offers mortality forecasts based on a spectrum of potential pandemic trajectories, determined by a parameter which varies between 0 and 1, yet remains uncalibrated. Meanwhile, \cite{zhou2021multi} introduced a tri-level model to simulate future mortality scenarios influenced by events similar to the COVID outbreak. The intricacies of the pandemic's progression are encapsulated within their model's third layer, which is heavily informed by expert insights. Further broadening the scope, \cite{chen2022modeling} developed a multi-country mortality framework which incorporates two distinct jump components: one signifying global pandemic shocks, and the other reflecting country-specific disturbances. \cite{robben2023catastrophe} applied a multi-population regime switching model to switch between periods of high volatility states (i.e. shock years) and low volatility states. However, they also assume uncorrelated one period shocks. \cite{richards2023} on the other hand discusses techniques for robust estimation of mortality rates in the presence of outliers, with a scenario included for gradually diminishing effects. 

All the previous literature that tries to account for vanishing jumps is using either expert opinion or simulation studies based on possible vanishing scenarios. Furthermore, the estimation procedure is based on a frequentist multi-step process. To overcome these issues and better reflect the dynamic nature of a pandemic, we present a modelling framework that allows for the inclusion of serial dependent shock components. As an example, we propose two ways of modelling the serial dependence, namely an autoregressive and a moving average type structure. For both in-sample and out-of-sample data, we show that our models outperform that of \cite{liu2015age}, where the shock components remain independent. In addition to single populations, we adopt our model to multi-populations as well and show that this leads to an even better in-sample fit. Lastly, we prove parameter identification for both frameworks. 
 
The remainder of this paper is organised as follows. Section \ref{modelling} provides the specification of our single-population model, which is composed of a baseline Lee-Carter model and a vanishing jump component with age-specific effect which can be autoregressive or moving average. Section \ref{estimation} details the estimation and procedure while section \ref{sec: Comparison} introduces the methods that we apply for the comparison and evaluation of our methods. Section \ref{applications} studies in-sample performance based on COVID-19 data from United States, Spain and Poland. In particular, we compare the performance of our model to the original Lee-Carter model and to the model of \cite{liu2015age} which does not include gradually vanishing effects. Section \ref{forecasting} discusses out-of-sample performance based on world wars data for England and Wales, both during times of war and normal times. Section \ref{sec:multipopulation} proposes a multi-population extension with vanishing jumps and compares single and multi-population models. Finally, Section \ref{conclusion} provides concluding remarks.

\section{Model specification}\label{modelling}

Our proposed model approach builds upon previous work on the Lee-Carter model and its extension with short-term jump effects by \cite{liu2015age}. Our model can be seen as a generalisation of their approach, allowing for more flexibility and accuracy in capturing pandemic effects on mortality rates. In this section, we start with a brief overview of both the Lee-Carter model and the Liu-Li model to establish the foundation for our proposed model.

When studying human mortality, the data at hand consist of death counts $D_{x, t}$ and central exposures $E_{x, t}$, where $x \in \{1, 2, \dots, A\}$ and $t\in \{1, 2, \dots, T\}$ represent a set of $A$ age groups and $T$ calendar years, respectively. We denote by $m_{x, t}$ the central death rate at age $x$ and calendar year $t$ given by 
\begin{equation*}
    m_{x, t}=\frac{D_{x, t}}{E_{x, t}}.
\end{equation*}

\subsection{The Lee-Carter model}

The Lee-Carter model \citep{lee1992modeling} is a well-known method for modelling mortality rates over time. It assumes that the logarithm of the central death rate $m_{x,t}$ for age group $x$ in year $t$ can be expressed as
$$
\ln \left(m_{x, t}\right)=\alpha_{x}+\beta_{x} \kappa_{t}+e_{x, t},
$$
where $\alpha_{x}$ represents the static level of mortality for age group $x$, $\kappa_{t}$ captures the variation of $\log$ mortality rates over time, $\beta_{x}$ measures the sensitivity of $\ln \left(m_{x, t}\right)$ to changes in $\kappa_{t}$, and $e_{x, t}\stackrel{\text{i.i.d.}}{\sim}\mathcal{N}\left(0, \sigma_{e}^{2}\right)$ is the error term. In the Lee-Carter model, $\alpha_{x}$ and $\beta_{x}$ represent age-specific effects, while $\kappa_{t}$ is a time-varying factor that captures the overall trend in mortality rates over time. Regarding the estimation, the frequentist approach is usually performed in two steps. First, parameters are obtained by maximising the model log-likelihood, then in a second step, projections are made by time-series techniques \citep{pitacco2009modelling}. In a Bayesian approach, the estimation and forecasting steps are performed in a single step, ensuring more consistent estimates in the estimation procedure \citep{cairns2011bayesian}. 

Despite its success in modelling mortality rates over time, the Lee-Carter model has a severe limitation when it comes to pandemics as the model assumes that mortality rates evolve smoothly over time, without sudden changes or shocks, driven by a simple random walk with drift. 

\subsection{The Liu-Li model}\label{s.LLM}

To introduce short-term jump effects, \cite{liu2015age} proposed an extension of the original Lee-Carter model, which includes an extra jump term as follows:
$$\ln \left(m_{x, t}\right)=\alpha_{x}+\beta_{x} \kappa_{t}+N_{t} J_{x, t}+e_{x, t},$$
Here, $\alpha_{x}, \beta_{x}$ and $e_{x, t}$ have the same meanings as in the original Lee-Carter model, while $\kappa_t$ is assumed to be a random walk with drift. Additionally, $N_{t}$ represents a binary random variable that equals one if a mortality jump occurs in year $t$ and zero otherwise. The authors assume that the $N_t$'s are i.i.d.~Bernoulli distributed  with parameter $p$, denoting the probability of a mortality jump in a calendar year. $J_{x, t}$ measures the effect of a mortality jump that occurred in year $t$ on age group $x$.

Three specific model variants were proposed, denoted as models J0-J1-J2. Model J1 is the closest to our model and is given by
\begin{equation}
    \ln \left(m_{x, t}\right)=\alpha_x+\beta_x \kappa_t+\beta_x^{(J)} N_t Y_t+e_{x, t},\label{modelJ1}
\end{equation}
where $Y_t$ denotes the effect or severity of the mortality jump at time~$t$. These jump effects are assumed to be i.i.d.~Gaussian variables. Compared to the Lee-Carter model, a new age pattern of pandemic effects $\beta_x^{(J)} $ is introduced. It is multiplied by the pandemic jump effect to capture age-specific variation, that is different from the period effects $\beta_x $. 

The model proposed by Liu and Li provides a valuable extension of the Lee-Carter model as it allows the inclusion of short jumps and different age patterns. However, it has two weaknesses: first, this model assumes that age patterns of different mortality shocks are the same, while historically, different age patterns have been observed. For example, half of the deaths caused by the 1918 flu pandemic occurred among 20- to 40-year-olds \citep{gagnon2013age}, while COVID-19 has affected mostly the most vulnerable people \citep{ferguson2020impact,o2021age}. In a Bayesian setting however, the pandemic effects $\beta_x^{(J)} $ are not considered to be fixed. The use of different priors provides a wide range of estimation possibilities. Second, the yearly jumps are independent and the Liu-Li model does not allow for a jump event lasting over several years with a vanishing effect as it can be observed for COVID-19. Our model presented in the next section addresses these shortcomings of the Liu-Li model.

\subsection{A new class of models allowing for serial dependent jump effects}

The limitations of the Lee-Carter model and its extension by \cite{liu2015age} motivate the need for a more flexible model that can capture the influence of a pandemic lasting over several years with a vanishing effect. Several recent studies have addressed this issue by proposing extensions to the Lee-Carter model. For instance, \cite{van2022estimating} and \cite{zhou2021multi} discussed the idea of a vanishing effect in the context of COVID-19, but did not try to estimate a corresponding parameter. To address this, we propose a model that allows for pandemic shocks that are transitory and vanishing over time. Our model, formulated in~\eqref{ourmodel_baseline}, extends the Lee-Carter model by adding a pandemic shock component, $J_t$, that captures serial dependence. 

The base line model is very similar to that of \cite{liu2015age} and given by 
\begin{equation}\label{ourmodel_baseline}
     \ln \left(m_{x, t}\right)= \alpha_x \,+\, \beta_x\, \kappa_t\,+\, \beta_x^{(J)} \,J_{t} \,+\,  e_{x, t}.
\end{equation}
Like \cite{liu2015age}, we model the time effect using a random walk representation 
\begin{equation}\label{eq: RWDrift}
     \kappa_t=\kappa_{t-1}+d+\xi_t,
\end{equation}
where $d\neq 0$ denotes the drift parameter and $(\xi_t)_t$ is a sequence of error terms. However, instead of setting $J_t = N_t \, Y_t$ as in Section~\ref{s.LLM}, where $Y_t$ denotes the magnitude of the jump effect and $N_t\in\{0,1\}$ indicates the jump occurrences, we introduce a more flexible approach that allows for a single shock to have an effect on consecutive years using ideas from time series models. 
We  propose two options to capture the serial dependence of the jump parameter $J_t$, namely an autoregressive structure in Subsection~\ref{sss.ar} and a moving average structure in Subsection~\ref{sss.ma}. We will refer to the former as the AR model and the latter as the MA model from hereafter. Of course, other structures to model serial dependence are possible. Both of our models have the advantage of modelling more complex patterns of mortality shocks and are easy to interpret. While the Liu-Li model only provides an estimate on the severity of a mortality shock, our model adaption provides an estimate of the severity, including how quickly the observed shock vanishes. This can help policymakers determine how long the effect of a particular mortality shock has lasted, and potentially adjust responses to future shocks.  

Additionally, the Bayesian framework we use for parameter estimation provides a full predictive distribution for all parameters, including $\beta_x^{(J)}$, which enables us to account for uncertainty. This flexibility allows us, for example, to accommodate different age patterns for different mortality jumps, that is for each future shock, we can obtain different realisations of the age patterns, which was one of the limitations of the Liu-Li model.

\subsubsection{Autoregressive structure}\label{sss.ar}
The discussions by \cite{van2022estimating} and \cite{zhou2021multi} suggest the use of an autoregressive type structure, where a pandemic effect slowly vanishes over time controlled by some parameter $a\in[0,1)$. This is also in line with Markovian epidemiological models for health states of COVID patients, see e.\,g.~\citet{bartolucci2021}. More precisely, we propose the following model:
\begin{equation}\label{ourmodel_AR}
\begin{split}
     \ln \left(m_{x, t}\right)&= \alpha_x \,+\, \beta_x\, \kappa_t\,+\, \beta_x^{(J)} \,J_{t} \,+\,  e_{x, t},\\
    J_{t}&= a\, J_{t-1} \,+ \, N_t \, Y_t.
\end{split}
\end{equation}

The parameters $Y_t$ and $N_t$ have the same meaning as in the model of \cite{liu2015age}. Though, the jump size $J_t$, which captures the impact of the pandemic shock on mortality rates, includes a vanishing effect controlled by the parameter $ a \in [0,1)$. The model \eqref{ourmodel_AR} is a (Bayesian) generalisation of the model J1 in Equation \eqref{modelJ1} of \cite{liu2015age}, that is recovered when $a=0$ and allows for a gradually vanishing effect when $a>0$. Indeed, if there is a mortality jump in year $t$, the impact on the log mortality rates is given by $\beta_x^{(J)} Y_t$ in year $t$, by $a\, \beta_x^{(J)} Y_t$ in year $t+1$, and so on. 

\subsubsection{Moving average structure}\label{sss.ma}

Instead of allowing the shock effect to slowly vanish over time we may also assume a moving average type structure with order $Q$. Here, the initial shock has an effect for a total of $Q$ periods and then disappears completely. 
A reasonable order $Q$, which indicates how long the initial shock lasts in the subsequent periods, has to be selected manually, a potentially challenging task. In addition, as the order $Q$ increases the number of parameters increases as well, which makes identification and estimation of parameters more complex. For sake of simplicity we therefore assume a MA(1) structure which can be defined as
\begin{equation}\label{ourmodel_MA}
\begin{split}
     \ln \left(m_{x, t}\right)&= \alpha_x \,+\, \beta_x\, \kappa_t\,+\, \beta_x^{(J)} \,J_{t} \,+\,  e_{x, t},\\
    J_{t}&= \, N_t \, Y_t + b\, N_{t-1} \, Y_{t-1}.
\end{split}
\end{equation}
Similar to the AR model, we assume that $b \in [0,1)$. Hence, the MA(1) model of \eqref{ourmodel_AR} can be seen as a generalisation of the \cite{liu2015age} model that is recovered if $b = 0$.

\section{Estimation procedure}\label{estimation}

For the estimation of mortality models, there are two common routes: either estimate the model on the central death rates directly (the traditional approach), or estimate the model on the first differences of the log mortality rates, also called mortality improvements. These two variants are referred to as \textit{Route I} and \textit{Route~II} in the terminology of \cite{haberman2012parametric}. In this paper, we proceed to Route II. It has the advantage of eliminating the static age effect $\alpha_x$ of the model, thereby reducing the number of identifiability constraints needed (see \cite{hunt2020identifiability} and the next Section \ref{identifiability}). We remark that \cite{mitchell2013} conducted an extensive study comparing the Route I and Route~II approach of multiple models in terms of in-sample fit, including the LC model as well as variants thereof, and found the Route II method to be superior. Moreover, \cite{wong2023bayesian} compared the in-sample fit of a LC model estimated on mortality rates with that of an age-period model estimated on mortality improvements rates and found the latter to be superior. In addition, \cite{mitchell2013} compared the width of the prediction intervals on held out data and found the ones of the Route~II approach too be narrower and more accurate in terms of coverage. 


In the Route II approach, the mortality improvements are modelled directly, defined as 
 \begin{equation}\label{eq: MortImp}
Z_{x, t}:=\ln \left(m_{x, t+1}\right)-\ln \left(m_{x, t}\right).
\end{equation}
Positive values of mortality improvement rates indicate worsening mortality conditions relative to the previous year, while negative mortality improvement rates display an improvement in mortality. Our model specification in~\eqref{ourmodel_baseline} can then be written as 
$$
Z_{x, t}=\beta_x\left(\kappa_{t+1}-\kappa_{t}\right)+\beta_x^{(J)}\left(J_{t+1}-J_{t}\right)+\varepsilon_{x, t},
$$
where $\varepsilon_{x, t}=e_{x, t+1}-e_{x, t}$. It follows, that $\varepsilon_{x, t} \stackrel{\text{i.i.d.}}{\sim} \mathcal{N}(0, \sigma_r^{2})$ with $\sigma_r^{2} = 2\sigma_e^{2}$. Using \eqref{eq: RWDrift}, this can be restated as
\begin{equation}
\begin{split}
        Z_{x, t}&=\beta_x\,\Delta \kappa_{t+1}+ \beta_x^{(J)}\,\Delta J_{t+1}+\varepsilon_{x, t}\\
       &=\beta_x\left(d+\xi_{t+1}\right)+\beta_x^{(J)}\,\Delta J_{t+1}+\varepsilon_{x, t},\label{ourmodel2} 
\end{split}
\end{equation}
 where $\Delta J_{t+1}=J_{t+1}-J_t$ and, similarly, $\Delta \kappa_{t+1}=\kappa_{t+1}-\kappa_t$.


\subsection{Identifiability constraints}\label{identifiability}

\noindent
Several mortality models, including the LC model, suffer from non-identifiability issues, meaning that different sets of parameters result in equivalent likelihoods and consequently the same fitted rates. In their paper, \cite{hunt2020identifiability} discuss the problem of non-identifiability in LC type models with multiple age or period functions at length and provide a general theorem for the selection of suitable constraints. Since our proposed model can be seen as an extension thereof, their logic can be applied to find the number of needed constraints. 

In a standard LC model, the age effect may be scaled and the time effect shifted to produce a new set of parameters resulting in the same fitted mortality rates. Thus, the parameters are not uniquely determined and can be transformed in two ways, namely
\begin{align}
\label{eq: LC Trans 1}
	\{{\tilde\alpha_{x}},\,{\tilde\beta_{x}},\,{\tilde\kappa_{t}}\} 
	&= \left\{\alpha_{x},\, \frac{1}{a}\,\beta_{x},\, a \,\kappa_{t}
	\right\} \, ,  \\ 
 \label{eq: LC Trans 2}
	 \{{\tilde\alpha_{x}},\,{\tilde\beta_{x}},\,{\tilde\kappa_{t}}\}
	 &= \left\{\alpha_{x}-b\,\beta_{x},\, \beta_{x}, \,\kappa_{t}+b	\right\} \,  ,
\end{align}

\noindent
for all $x \in \{1, 2, \dots, A\}$ and $t\in \{1, 2, \dots, T\}$.

In principle, the same problem holds for our model formulation as well. 
Using matrix notation we can rewrite the first equation of~\eqref{ourmodel_baseline} in a compact way.  Let $\boldsymbol{B}_x = (\beta_x,\beta_x^{(J)})^{\intercal}$ and $\boldsymbol{K}_t = (\kappa_t, J_t)^{\intercal}$, then 
\begin{equation}\label{eq: OwnMod_inMatrix}
    \ln (m_{x,t}) =\alpha_x+ \boldsymbol{B}_x^{\intercal} \boldsymbol{K}_t + e_{x,t}.
\end{equation}
The model in \eqref{eq: OwnMod_inMatrix} has the same structure as the classical LC model and can be thought of a multivariate extension, coined LC2 in the terminology of \cite{hunt2020identifiability}. Unsurprisingly, the model in~\eqref{eq: OwnMod_inMatrix} suffers from non-identifiability. Let there be a  matrix $\mathbf{A} \in \mathbb{R}^{2 \times 2}$ that is invertible and a matrix $\mathbf{D} \in \mathbb{R}^{2 \times 1}$. Then, according to \citet[Theorem 1]{hunt2020identifiability} equations \eqref{eq: LC Trans 1} and \eqref{eq: LC Trans 2} can be generalised to higher dimensions where the parameters of \eqref{eq: OwnMod_inMatrix} can be transformed using


\begin{align}
    \label{eq: LC2 Trans 1}
	\{\tilde{\alpha},\tilde{\boldsymbol{B}}_x,\tilde{\boldsymbol{K}}_t\} 
	&= \left\{\alpha_x, \mathbf{A}^{-1}\boldsymbol{B}_x, \mathbf{A}\boldsymbol{K}_t
	\right\} \\
 \label{eq: LC2 Trans 2}
	\{ \tilde{\alpha},\tilde{\boldsymbol{B}}_x,\tilde{\boldsymbol{K}}_t\} 
	&= \left\{\alpha_{x}-\mathbf{D}^{\intercal}\boldsymbol{B}_x, \boldsymbol{B}_x, \boldsymbol{K}_t+\mathbf{D}	\right\}.
\end{align}

Since matrix $\mathbf{A}$ is $ (2 \times 2)$ and $ \mathbf{D}$ is $(2 \times 1) $, there are in total six free parameters meaning that we have to impose six identifiability constraints for the model in~\eqref{eq: OwnMod_inMatrix}. 

However, note that Equation~\eqref{eq: OwnMod_inMatrix} includes an age specific intercept. When differencing the log mortality rates to obtain $Z_{x,t}$, i.e. applying the Route~II estimation method, the age specific intercept cancels. In this case, we obtain that $\mathbf D=0_{2\times 1}$ in \eqref{eq: LC2 Trans 2} and no further identifiability issues arise from \eqref{eq: LC2 Trans 2}.  Consequently, there is a reduced set of identifiability constraints, namely the four entries of the matrix~$\mathbf A$, as only transformations of \eqref{eq: LC2 Trans 1} are relevant \citep[Appendix A.]{hunt2020identifiability}. Hence, by applying the Route~II estimation approach we can reduce the amount of identifiability constraints needed from six to four, by cancelling out the static age function $\alpha_{x}$ due to differentiation of the death rates.  

As we prove in the Appendix, identification is ensured by imposing the standard sum-to-one constraints on age parameters, that is 
\begin{equation*}
    \sum_{x=1}^{A} \beta_{x} = 1 \quad \text{ and } \quad  \sum_{x=1}^{A} \beta_{x}^{(J)} = 1,
\end{equation*}
and by using  corner constraints on the first differenced time parameters, $\Delta J_{2} = 0$ and $\xi_2 = 0$, resulting in a total of four constraints. This is enough to identify the drift~$d$ and the parameters of the mortality improvement rates $Z_{x,t}$, i.e.~$\beta_x,\,\beta_x^{(J)}, \,\Delta \kappa_t$, and $\Delta J_t$. If we assume additionally that $J_1 = J_2 = 0$, then we can identify all of $(J_t)_t$ iteratively using $(\Delta J_t)_t$. 

However, depending on the model structure assumed for $J_t$, additional constraints need to be imposed to identify the jump occurrences $N_t$ and either the autoregressive parameter~$a$ or the moving average parameter~$b$. 
In particular, we assume knowledge of a known time point $\tilde{t}$ after a first shock event where there is no jump, that is, $N_{\tilde{t}} = 0$. For example, this can be set to $\tilde{t} = T$. For the MA setting, some further constraints are needed. Details can be found in the Appendix~\ref{sec: ProofIdentification}.

It should be noted that these are not the only identification constraints that can be set. Given the recommendation by \cite{hunt2020identifiability}, another possibility is to adopt a true normalisation scheme for the age parameters. That is, instead of a sum-to-one constraint, we could set the age parameters to have an Euclidean norm of one, that is
\begin{equation*}
    \norm{\beta_{x}}_{2}^2 = 
    \sum_{x=1}^{A} \left( \beta_{x} \right)^{2} = 1 \quad \text{ and } \quad  
     \norm{\beta_{x}^{(J)}}_{2}^2 = 
    \sum_{x=1}^{A} \left( \beta_{x}^{(J)} \right)^{2} = 1.
\end{equation*}
The above identification can be achieved, for example, using $\mathbf{QR}$ decomposition, which results in two orthonormal age vectors that do not only have a norm of one, but are also orthogonal to each other, resulting in a dot product of zero, that is $\sum_{x} \beta_x \beta_x^{(J)} = 0$. When adopting the identification scheme using $\mathbf{QR}$, the corner constraint on the time dependent parameters needs only to be set on the jump effect, thus $\Delta J_{2}  = 0$.  

Both identification schemes have been successfully implemented and give unique parameter estimates. For sake of model comparison, we choose to go with the standard sum-to-one constraints, as these are the ones selected by \cite{liu2015age}.

\subsection{Priors}

To estimate the parameters in Equation \eqref{ourmodel2}, we consider a Bayesian approach to inference. It is based on the idea of updating prior beliefs with the data at hand to obtain a posterior distribution of the parameters. For the selection of priors there are many options available. If not stated otherwise, we employ the use of so called weakly informative priors. Here, the prior should rule out unreasonable values but not be too restrictive so that it rules out plausible values. 

As stated in Equation~\eqref{eq: RWDrift}, we assume that the time-dependent parameter $\kappa_t$ follows a random walk type representation, which we model using a normal prior: $\Delta\kappa_t \stackrel{i.i.d.}{\sim} \mathcal{N}(d,\xi_t)$. The jump effects are given a half normal prior, such that, $Y_t \stackrel{i.i.d.}{\sim} \mathcal{N}^{+}(\mu_Y,\sigma_Y)$, because the focus on this paper is on catastrophic mortality jumps. This is in contrast to the approach of \citet{liu2015age}, where $Y_t$ is assumed to be Gaussian to preserve the tractability of the likelihood. However, they note that other distributions might produce a better fit. The jump occurrence is modelled using a Bernoulli distribution $N_t \stackrel{i.i.d.}{\sim} \operatorname{Bern}\left(p\right)$. The age-specific parameters $\left(\beta_{1},\dots,\beta_{A}\right)$ and $\left(\beta^{(J)}_{1},\dots,\beta^{(J)}_{A}\right)$ are given multivariate Dirichlet priors, which implicitly impose the sum-to-one constraints. For both the autoregressive parameter $a$ as well as the moving average parameter~$b$ we assume a slightly informative normal prior truncated from zero to one with mean of zero and standard deviation of 0.4. This parameterisation favours smaller values of $a$ respectively $b$ with the most prior mass around zero. Hence, there needs to be evidence by the likelihood to move the posterior estimate away from zero.

For the hyperparameters we choose a mix between weakly informative and informative priors. Starting with the drift parameter $d$ we assume a normal prior with a smaller standard deviation. The normal prior guarantees that $\text{Pr}(d\neq 0)=1$, as required for identification of our parameters. For the jump probability we impose a rather informative hyperprior, where $p \sim \text{Beta}(1,20)$, which strongly favours small values of $p$. This is for the following reason. The parameter $J_t$ is intended to model extreme events, not just noise. Thus, a shock should be something that occurs rarely, less than every few years, rather than some regular ups and downs. The latter type of effects should be captured by the noise term of the random walk coefficient and not considered a shock. Thus we know in advance that mortality shocks appear infrequently and that the jump proportion has to be low. This information can be included in the prior. Moreover, when experimenting with uninformative hyperprior settings, e.g. the well-known Jeffery's prior $p \sim \text{Beta}(0.5,0.5)$, we noticed the tendency of the model to flag many more smaller bumps as potential shocks which sometimes led to a lack of convergence of the age-specific jump parameters $\beta_{x}^{(J)}$. This effect can be explained by the fact that our data contains too few shock events to calibrate reasonably well without any prior information. Moreover, several papers in the literature rely on expert knowledge to handle this issue. Imposing an informative prior on $p$ alleviated the problem in a data driven way based on the comparatively mild information of low shock frequency and helped with convergence. However, we note that our specific choice of $p$ is of course subjective. We could have used a different parameterisation, e.g. $p \sim \text{Beta}(1,10)$ or $p \sim \text{Beta}(1,15)$. A small sensitivity analysis showed that either hyperprior led to a similar posterior.

Using our modelling approach, we want to capture shocks that have a positive (i.e. increasing) effect on death rates. We therefore assume that the jump effect $Y_t$ can only take on positive values and impose a half normal prior, which assumes a positive mean. Therefore, we choose a half-normal prior on the jump mean parameter $\mu_{Y}$ as well. Lastly, all standard deviations are given half-normal priors. An overview on specific values for the hyperparameters can be found in the Appendix~\ref{S95}.

\subsection{Parameter estimation}

To estimate the parameters of our proposed model, we use NIMBLE \citep{nimble-article:2017}, a system for programming statistical algorithms in R. NIMBLE provides a flexible and intuitive framework for model specification while supporting programming functions that adapt to model structures. Moreover, it allows for the selection of multiple samplers that include the well known MCMC methods as well as Hamiltonian Monte Carlo (HMC). For each parameter a different sampler can be chosen allowing for great flexibility and efficient computation. NIMBLE can be accessed via the \emph{nimble} package in R \citep{nimble-software:2022}. To be able to use the HMC sampler, the package \emph{nimbleHMC} \citep{nimbleHMC:2024} must be downloaded as well. If not stated otherwise, NIMBLE uses a conjugate Gibbs sampler where possible as well as Metropolis-Hastings. However, the latter tends to be very inefficient due to the high autocorrelation of the samples. As a result, we choose to change the samplers of multiple variables resulting in improved mixing performance. Moreover, to be able to use the flexibility of NIMBLE, we have chosen to parameterise the Dirichlet in terms of a normalised Gamma distribution, which allows us to select from a greater pool of available samplers, like the multivariate slice sampler of \cite{tibbits2014} for example or an HMC sampler. A justification of the construction of a Dirichlet using the Gamma distribution can be found in the Appendix~\ref{sec: DirichAsGamma}). In addition, for the jump occurrence $N_t$, NIMBLE uses a special Gibbs sampler for binary-valued variables. A list can of the specific samplers for each parameter be found in the Appendix~\ref{S96}. 

To assess the convergence of all model parameters, we employ three widely recognised diagnostics: the split-$\hat{R}$ statistic as well as variants of the effective sample size, namely bulk effective sample size (Bulk-ESS) and tail effective sample size (Tail-ESS). These diagnostics are implemented using the \textit{rstan} package \citep{rstan}, and for a more comprehensive discussion on their use, we refer to \cite{vehtari2021rank}.

In our analysis, we utilise two chains for each country and model. We discard the initial 7,500 iterations of each chain as ``burn-in", ensuring that the chains have stabilised. Subsequently, we draw an additional 10,000 samples per chain, with only every 10th sample being retained for inference. For a comprehensive understanding of parameter convergence and additional details regarding the MCMC settings, please refer to the Appendix. Even without parallelisation, the total run time of the model is rather short and took around 5 minutes on a Intel i5-8365U CPU @ 1.60GHz processor with 16,0 GB RAM. 

\section{Model comparison} \label{sec: Comparison}
After having estimated the parameters we want to assess both the in as well as out-of-sample fit of our models. 

\subsection{In-sample comparison}

To assess the in-sample fit of the models in question, we calculate the widely applicable or Watanabe-Akaike Information Criterion \citep[WAIC;][]{watanabe2010asymptotic}. A lower WAIC value indicates a better-fitting model among the alternatives, as is the case with most information criteria. What distinguishes WAIC is its fully Bayesian nature, since it considers the entire posterior distribution for model evaluation. In 
addition to WAIC, the in-sample fit may be compared using cross validation. 

Consider some data, $\mathbf{y}=(y_{1},\dots, y_{N}) $, which is modelled as independent given the parameter $\theta$, hence $p(\mathbf{y}|\theta) = \prod_{i=1}^{N}p(y_i|\theta)$. A standard quantity in Bayesian analysis is the log predictive density (lpd), 
\begin{equation*}
		\text{lpd} = \sum_{i=1}^{N} \log p(y_i|\mathbf{y}) = 
  \sum_{i=1}^{N} \log \int p(y_{i}|\theta) p(\theta|\mathbf{y}) d\theta.
\end{equation*} 
Using the lpd, we can calculate the WAIC by
\begin{align*}
		\text{WAIC} &= -2\text{lpd} + 2p_{waic}\\
		p_{waic} &= \sum_{i=1}^{N} \text{Var}\left( \log p(y_i|\theta) \right).
\end{align*}
To calculate the WAIC in practice, the lpd and $p_{waic}$ have to be estimated using posterior draws. Let $ \theta^{(s)} $ denote the $s$-th sample from the posterior, with $ s = 1,\dots S $. The inner expectation of the lpd can be approximated by:
\begin{equation*}
		\widehat{\text{lpd}} = \sum_{i = 1}^{N} \log \left( \frac{1}{S} \sum_{s=1}^{S} p(y_{i}|\theta^{(s)})\right).
\end{equation*}
An estimate of $p_{waic}$ is given by
\begin{equation*}
    \widehat{p}_{waic}  = \sum_{l = 1}^{n} V_{s = 1}^{S} \log p(y_{l}|\theta^{(s)}),
\end{equation*}
where $V_{s = 1}^{S}$ represents the sample variance \citep{vehtari2017practical}. 

The Bayesian leave-one-out cross validation (LOO-CV) is based on the log predictive density given the data without the $i$-th data point $p(y_{i}|y_{-i})$. In practice, it is calculated as 
\begin{equation*}
		\text{lpd}_{\text{loo}} = \sum_{i = 1}^{N} \log p(y_{i}|y_{-i}),
\end{equation*}
where 
\begin{equation*}
		p(y_{i}|y_{-i}) = \int p(y_{i}|\theta) p(\theta|y_{-i}) d\theta
\end{equation*}
denotes the leave-one-out predictive density without the $i$-th data point. In the above setting, exact cross validation would require refitting the model $N$ times. However, $p(y_{i}|y_{-i})$ can be approximated using importance sampling \citep{vehtari2017practical}. A model with a higher $\text{lpd}_{\text{loo}}$ indicates a better model fit by superior predictive performance. Oftentimes, $\text{lpd}_{\text{loo}}$ is provided on the deviance scale, that is $ \text{LOO-CV} = -2\text{lpd}_{\text{loo}} $, where a lower score suggests the better fit. Both the calculation of the WAIC and LOO-CV is implemented in the \emph{loo} package in R \citep{loo2023}.

\subsection{Mortality forecasts}\label{ch: MortForecasts}

As our model estimates mortality improvement rates, the generated forecasts are of the same form. However, to obtain forecasts of future (log) death rates, the mortality improvement rates can be transformed. Using Equation~\eqref{eq: MortImp}, we can calculate the death rate at time $t+1$ as follows:
\begin{equation}\label{eq: Projection}
\ln\left(m_{x,t+1}\right) = \ln\left(m_{x,t}\right) + Z_{x,t}.
\end{equation}
Let the projection periods be denoted as $h \in \{1, \dots, H\}$. To generate a $h$-step ahead forecast of $\ln\left(\hat{m}_{x,{T+h}}\right)$, we proceed by generating new values from our posterior predictive distribution for \sloppy  $ Z_{x,{T+1}}, \dots, Z_{x,{T+H}} $ and apply Equation~\eqref{eq: Projection} recursively. This procedure can be repeated $S$ times to obtain $S$ draws from the posterior predictive distributions of $\ln\left(\hat{m}_{x,{T+h}}\right)$. We can then derive prediction intervals using Monte Carlo simulations.

To make predictions for future values of $Z_{x,{T+1}}, \dots, Z_{x,{T+H}}$, we must also generate new values for time-dependent parameters. For example, to predict $Z_{x,{T+1}}$ for the model with the autoregressive structure of \eqref{ourmodel_AR}, we can follow these steps for each posterior draw ($s = 1, \dots, S$):
\begin{enumerate}[leftmargin=4em]
    \item[\emph{Step 1: }] Generate new values of $N^{(s)}_{{T+1}}$ by first drawing a value of $p^{(s)}$ from the posterior distribution and then sampling $N^{(s)}_{{T+1}}$ from a Bernoulli distribution with parameter $p^{(s)}$.
    \item[\emph{Step 2: }] Generate new values of $J_{{T+1}}^{(s)}$. Start by drawing $\mu_{Y}^{(s)}$ and $\sigma_{Y}^{(s)}$ from the posterior distribution. Then sample a new value of $Y_{{T+1}}^{(s)}$ from a normal distribution with mean $\mu_Y^{(s)}$ and standard deviation $\sigma_Y^{(s)}$. Afterwards, draw $a^{(s)}$ and $J_{T}^{(s)}$ from the posterior distribution. Use the newly generated $N^{(s)}_{{T+1}}$ from Step 1 to compute a future value of $J_{{T+1}}^{(s)}$.
    \item[\emph{Step 3: }]Generate a new error term $\varepsilon_{x,{T+1}}^{(s)}$ by sampling from a normal distribution with mean 0 and standard deviation~$\sigma_r^{(s)}$.
    \item[\emph{Step 4: }] Obtain the $s$-th posterior draw for the remaining parameters and substitute all values into Equation~\eqref{ourmodel2} to generate $Z^{(s)}_{x,{T+1}}$.  
\end{enumerate}

\noindent
These steps are then repeated for $T+1, \dots, T+H$ to generate future log death rates: $\ln\left(m_{x,t+1}\right),\dots, \ln\left(m_{x,t+H}\right)$.

\subsection{Out-of-sample comparison}
In addition to the in-sample comparison we can compare the predictive accuracy of our models on out-of-sample data. Since we are estimating the parameters in a Bayesian setting, our forecasts denote an entire predictive distribution rather than a single point, i.e. a mean forecast. Scoring rules provide a means to compare the accuracy of a predictive distribution of competing models around an observed data point. An overview on the idea as well as examples of scoring rules can be found in \cite{gneiting2007a}. Scoring rules are similar to information criteria in the sense that a lower score denotes a better fit. Two popular examples of scoring rules are the negative log score and the continuous ranked probability score (CRPS), which is more robust. 

Suppose that the data for $T$ years is split into training and validation set, where $Z_{1:M}$ denotes the mortality improvement rates for all ages of the first $M$ years used to fit the model with corresponding parameters $\boldsymbol{\theta}_M$. 
The log score (LogS) is given by the negative logarithm of the predictive density for a future $h$-step ahead observation and defined as 
\begin{equation}\label{eq: logScore}
    \text{LogS}(Z_{x,M+h}) = - \log p(Z_{x,M+h}|Z_{1:M}) = - \log \int p(Z_{x,M+h}|\boldsymbol{\theta}_M) p(\boldsymbol{\theta}_M | Z_{1:M} ) d \boldsymbol{\theta}_M.
\end{equation}
The CRPS on the other hand is defined in terms of predictive CDF given by
\begin{equation} \label{eq: CRPS}
    \text{CRPS}_{x,M+h} = \int \left[ F(z|Z_{1:M}) - \mathbf{1}_{Z_{x,M+h}\leq z}
    \right]^2 dz,
\end{equation}
where $\mathbf{1}$ denotes the indicator function and $F(z|Z_{1:M})$ the CDF of the predictive density $p(z|Z_{1:M})$. Evaluation of \eqref{eq: logScore} and \eqref{eq: CRPS} requires replacing the PDF and CDF of the predictive density with their empirical counterparts obtained using samples from the posterior. Both scoring rules are available in the R package \emph{scoringRules} \citep{jordan2019}. When deciding between two competing models, the one with the lower total score summed over all future observations and age groups is considered the better choice. 

In addition to scoring rules, we also compare the mean squared error (MSE) and mean absolute error (MAE) of our posterior mean forecasts of the log mortality rates using Equation \eqref{eq: Projection} and the observed log mortality rates.



\section{Data analysis during COVID-19: In-sample performance}\label{applications}

Our primary objective is to introduce an enhanced Lee-Carter model capable of accurately capturing the fluctuations in log death rates driven by the COVID-19 pandemic. The classical approach of assessing predictive accuracy through data splitting (training and testing data via out-of-sample validation) faces unique challenges in our context. The pandemic predominantly impacts the most recent years of data, making it impractical to exclude these years and estimate parameters using only the earlier data. Such an approach would overlook the pandemic's specific dynamics. 

Given these challenges, our analysis of the COVID data focuses solely on evaluating the in-sample fit of the models. To demonstrate the effectiveness of our refined Lee-Carter approach, we will showcase its performance in three distinct countries: the United States, Spain, and Poland. These countries were chosen as illustrative examples due to their significant experiences with the COVID-19 pandemic and the availability of relevant mortality data. In the following sections, we will delve into the model's parameter estimates and its capacity to capture the unique mortality patterns observed during the pandemic in each of these nations. 

The data used for this study was mainly sourced from the Human Mortality Database (HMD)\footnote{HMD website: \url{https://www.mortality.org/}} and Eurostat\footnote{Eurostat website: \url{https://ec.europa.eu/eurostat/de/web/main/data/database}}. We focused on three western countries that experienced significant COVID-19 impacts, as indicated by deaths per 100,000 population\footnote{COVID-19 data: \url{https://coronavirus.jhu.edu/data/mortality}}.  Specifically, we analysed unisex populations from the United States (US), Spain and Poland. 

For Spain and Poland, we obtained data from Eurostat, however, for Poland, data was available starting in 1990. To create a consistent dataset we selected this to be the starting year for all countries in question. We obtained yearly counts of death from Eurostat up until 2022 and combined them with provisional counts of weekly deaths for 2023, aggregating the latter to obtain annual death counts. Exposure was available from Eurostat until 2023 for Spain, however only until 2020 for Poland. Population estimates of the latter for 2021 - 2023 were obtained from Statistics Poland 
\footnote{Statistics Poland: \url{https://stat.gov.pl/en/topics/population/population/}}. In the case of the US, we acquired death and exposure counts up to 2021 from HMD. For the years 2022 and 2023 we obtained provisional counts of deaths and exposure from the National Center for Health Statistics (NCHS) and provisional estimates of the mid year population from the United States Census Bureau. Both were available to download from the Centers for Disease Control and Prevention (CDC) website\footnote{CDC data: \url{https://wonder.cdc.gov/mcd-icd10-provisional.html}}. To be able to combine multiple data sets by country we had to go with the lowest granularity regarding age group size. Provisional counts of deaths for US were given for age groups of width 10, except for the youngest age group. Hence, for sake of comparison, we choose to adopt this as the general age structure for all countries. Meaning that the data was organised into a total of $A = 10$ age groups from `$<5$', `$5$-$14$' up until `$85+$'. 

A summary of the data sources can be found in Table~\ref{tb: Data Sources}.

\begin{table}[ht]
   \begin{center}
    \begin{minipage}{\textwidth}
    \caption{Sources of mortality data}\label{tb: Data Sources}
	\begin{tabularx}{\linewidth}{XXl}
		\toprule
		Country & Year  & Source    \\
		\midrule
		\textbf{Counts of Death} & & \\
        Poland & 1990 - 2023 & Eurostat \\
        Spain & 1990 - 2023 & Eurostat  \\
        United States & 1990 - 2021 & HMD \\
        &               2022 - 2023    &CDC \\ 

        \vspace{0.05cm}
        \textbf{Population Estimate} & & \\
        Poland & 1990 - 2020 & Eurostat \\
        &        2021 - 2023 & Statisics Poland\\
        Spain & 1990 - 2023 & Eurostat \\
        United States & 1990 - 2021 & HMD \\
        &               2022 - 2023    & US Census Bureau\\ 
        \bottomrule
	\end{tabularx} 
     \end{minipage}
\end{center}
\end{table} 

For the COVID data, we assume $N_T = 0$ for the US, Spain and Poland in order to ensure  identifiability of the parameters for our AR and MA model.

\subsection{United States}
We applied our models to the US mortality data spanning from 1991 to 2023, comparing it with the Liu-Li model. When evaluating the goodness of fit using the WAIC and LOO-CV metric, our proposed models demonstrate superior performance (see Table \ref{tb: WAIC_US}). More precisely, the MA offers the best in-sample fit, according to our metrics, followed by the AR model. However, the difference between all three models is moderate, which is not unexpected given that the models diverge mainly over a period of three years after the start of the COVID pandemic, i.e. 2021-2023. 

\begin{table}[ht]
   \begin{center}
    \begin{minipage}{\textwidth}
			\caption{In-sample fit comparison of the Liu-Li and own models on COVID-19 data for multiple countries. Bold value denotes best of the column.}
			\label{tb: WAIC_US}
			\begin{tabularx}{\textwidth}{XYYY}
				\toprule
				Model& US & Spain & Poland  \\ 
				\midrule
				\textbf{WAIC} & & & \\ 
				AR & -1463.47 & -1193.73 & -1174.92 \\ 
                MA & \textbf{-1465.79} & \textbf{-1194.63} & \textbf{-1181.99} \\ 
                Liu-Li & -1462.45 & -1187.28 & -1178.66 \\ 
                \vspace{0.1cm}
				\textbf{LOO-CV} & & & \\
				AR & -1455.42 & -1188.21 & -1166.10 \\ 
                MA & \textbf{-1458.00} & \textbf{-1188.66} & \textbf{-1176.04} \\ 
                Liu-Li & -1453.39 & -1180.85 & -1170.29 \\ 
				\bottomrule
			\end{tabularx}
   \end{minipage}
   \end{center}
\end{table}

\begin{figure}[ht]
	\centering
	\includegraphics[height=8.5cm]{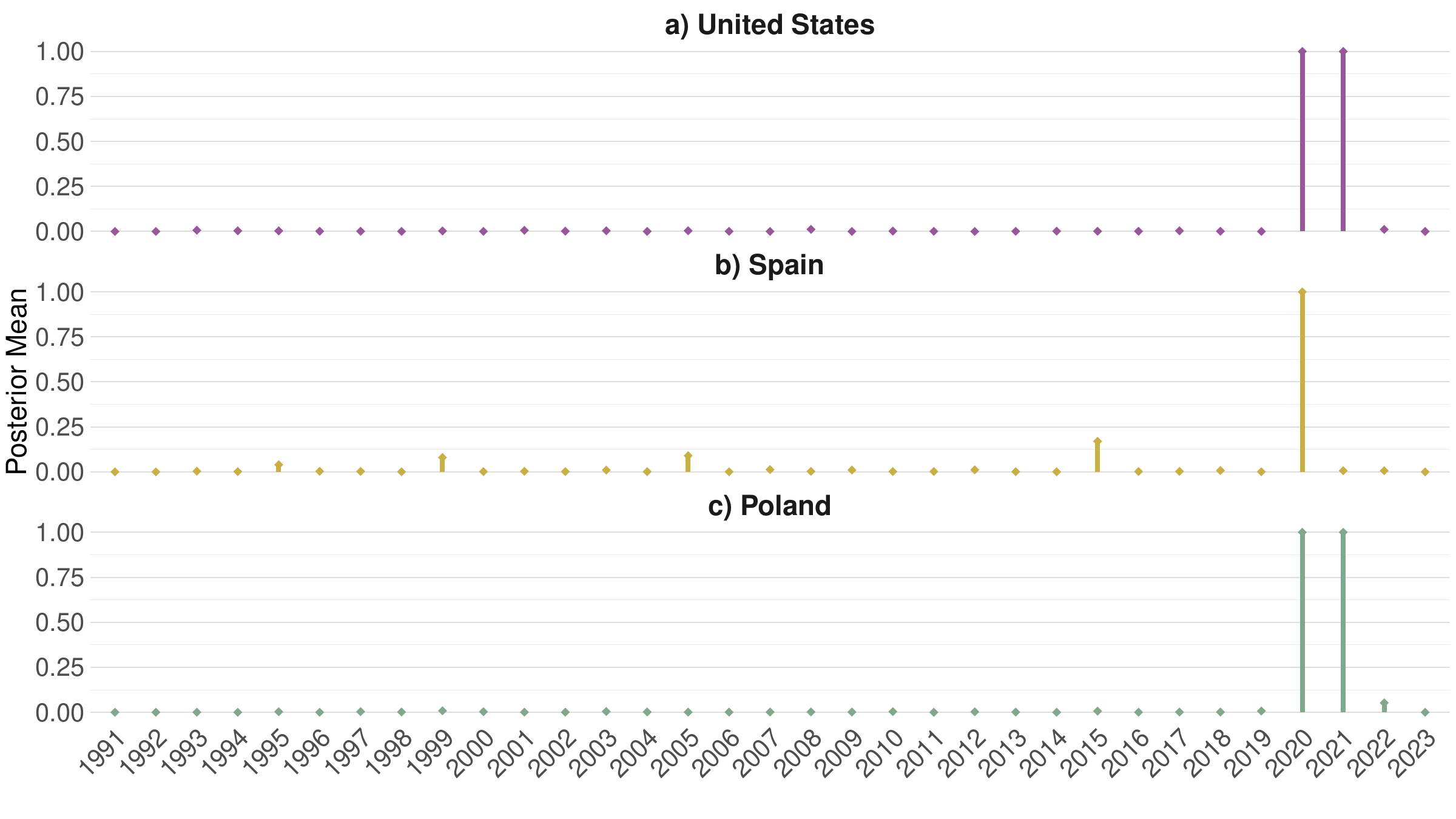}
	\caption{Comparison of posterior mean estimates for $N_t$ across time for all countries.}
	\label{img: JumpOccurenceComp}
\end{figure}

Looking at the parameter estimates of the MA model in more detail, we notice that the model demonstrates remarkable confidence in its assessments of the data for 2020 and 2021, with posterior mean estimates of $N_{2020}$ and $N_{2021}$ equalling 1 and a posterior standard deviation of 0.  
This high level of confidence indicates that the model considered these years as shock years with absolute certainty. Conversely, the other years showed extremely low to negligible posterior means, making them unsuitable candidates for jump years in the model. It should be noted, that this denotes one of the advantages of using a Bayesian approach to parameter estimation, in that $N_t$ is treated as a parameter rather than a random variable, allowing us to explicitly analyse the estimated timing of a mortality shock. For an overview on the estimated occurrences of jump years see panel a) of Figure \ref{img: JumpOccurenceComp}. Another set of interesting parameters denote those of the jump effect $J_t$, namely $\mu_Y$ and $\sigma_Y$ and $b$ (see Figures~\ref{img: JumpParamComp} and \ref{img: VanishingParam}). In the MA model these parameters yield posterior mean estimates, with values of 1.15 (80\%-PI [0.24,\,2.02]) and 1.27 (80\%-PI [0.37,\,2.47]), respectively. The moving average parameter $b$ is estimated to be medium sized with a posterior mean of 0.5 (80\%-PI [0.33,\,0.68]). It's worth acknowledging the relatively large standard deviation in the estimates. This variability is not unexpected, as the model predominantly considers only two years extreme events, meaning that parameter estimation is based on these two years only. Furthermore, in such data-scarce scenarios, the prior distribution significantly influences posterior estimates, especially via the choice of hyperparameters. 

As an example for a single country, we can compare the posterior estimates of 
$b$ with that of the AR parameter, $a$. Here we see a slightly lower  posterior mean for $a$ of 0.39 (80\%-PI [0.30,\,0.48]). Intuitively, this makes sense, since in the AR model the jump parameter $J_t$ is given as a linear combination of all of the past jump effects and not just the immediate past. However, regardless of the model choice, the estimated parameters $a$ and $b$ suggest a substantial amount of serial dependence in the data that is not captured by the Liu-Li approach, as indicated by the probability of $b$ and $a$ being greater than 0.1 being 99.6\% and 99.7\% respectively, and their maximum a posteriori (MAP) values being 0.48 and 0.41 respectively. All posterior estimates of the AR and MA models can be found in Table~\ref{tb: Est_US_AR} and Table~\ref{tb: Est_US_MA} in the appendix. 


The posterior mean estimates of $\mu_Y$ and $\sigma_Y$ underscore the substantial impact of the COVID-19 pandemic on mortality rates in the US. However, this influence appears to be more diffuse, affecting a broader age spectrum rather than concentrating on specific age groups. Intriguingly, the mortality jump pattern, as represented by $\beta_x^{(J)}$, exhibits a plateau in the middle age range (from 15-24 to 45-54), with a milder impact observed at higher ages. This pattern aligns with findings from \cite{faust2022two}, which revealed the most significant relative increase in mortality among the 18-49 age group, corresponding to the working population that played a central role in the spread of COVID-19 \citep{monod2021age}. The posterior mean estimates along with its 80\%-PI of $\beta_x^{(J)}$ are depicted in Figure \ref{img: AgeMortPattern} across all age groups. 




\subsection{Spain}

Comparing the model fit of Liu-Li and our models using the WAIC and LOO-CV, we again see a superior performance of the models that allow for a serial dependent jump component (cf. Table \ref{tb: WAIC_US}). The WAIC of both the AR and the MA model are very comparable, however the MA model has the lower estimate.  

Both the Liu-Li as well as the two other models estimate the posterior probability of a jump occurring in 2020, i.e. $N_{2020}$ to be one, while most of the other years, including 2021, have a negligible posterior mean estimate close to zero. However, the effect of the pandemic does not disappear in 2021 as the vanishing effect~$b$ has a posterior mean of 0.21 (80\%-PI [0.1,\,0.32]), indicating a medium sized effect (cf Figure~\ref{img: VanishingParam}). This means, that that on average 22\% of the shock in 2020 is still present in 2021. 

\begin{figure}[ht]
	\centering
	\includegraphics[height=8.5cm]{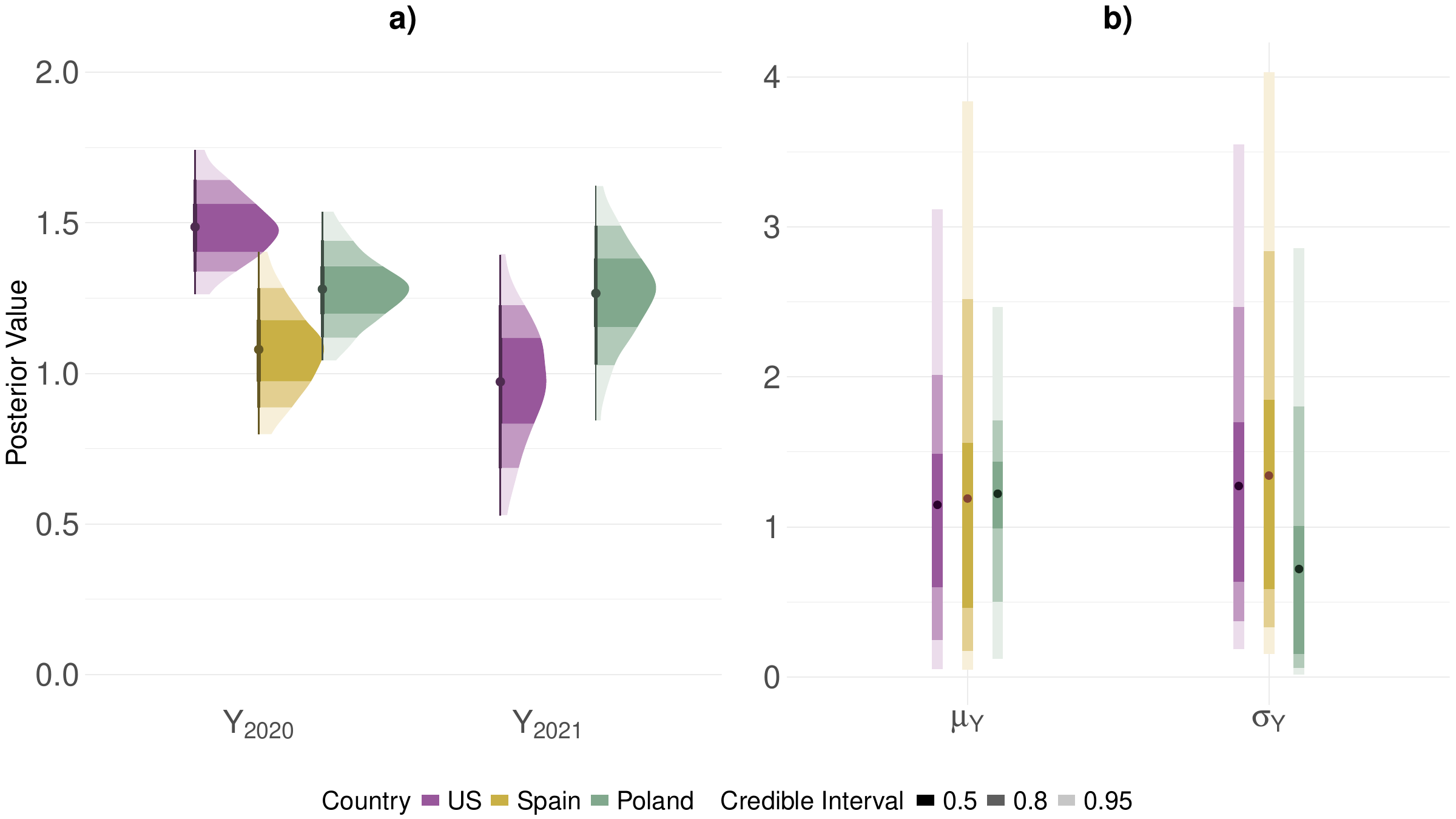}
	\caption{Comparison of posterior distribution for the jump parameters $\mu_Y$ and $\sigma_Y$ across countries. The point plotted below or within the density represents the posterior mean.}
	\label{img: JumpParamComp}
\end{figure}

The severity of the COVID effect in Spain is on a similar level to that of the US, with a posterior mean estimate of $\mu_{Y}$ given by 1.19 (80\%-PI[0.17,\,2.52]), while that of~$\sigma_{Y}$ is 1.34 (80\%-PI[0.33,\,2.84]). Posterior distributions can be found in panel b) of Figure~\ref{img: JumpParamComp}. Looking at the age pattern of the posterior shock, it is clear that the pandemic had a greater impact on older age groups than on younger ones. Two notable things are seen for the posterior estimates of $\beta_x^{(J)}$ depicted in Figure~\ref{img: AgeMortPattern}. First, there is a small plateau in the posterior mean in middle ages. At higher ages the posterior mean then decreases, only to increase again at the end of the age spectrum, reaching its global maximum in the 75-84 age group. Thus, both the mortality rates of the medium aged as well as the elderly were affected most by the COVID-19 pandemic. Posterior estimates of all parameters including uncertainty quantification for both the AR and MA model can be found in Table~\ref{tb: Est_Sp_AR} and \ref{tb: Est_Sp_MA} in the Appendix.

\subsection{Poland}

Comparing the model fit of all three models using the WAIC and LOO-CV, we again see superior performance by a model that captures the serial dependence of the COVID-19 pandemic (c.f. Table \ref{tb: WAIC_US}). However, for Poland only the MA model does better than the Liu-Li approach in terms of in-sample fit. This can be explained by the following observation in that the COVID-19 pandemic still has a substantial effect in 2022 only to disappear completely in 2023. The specific structure of the MA model allows such a pattern to be captured, while in the AR the COVID shock still affects mortality rates in 2023. The difference in WAIC is thus explained by the difference in model fit for 2023. Looking at the WAIC estimates for 2023 only, we see very comparable values for the Liu-Li and MA models, while that of the AR model is significantly lower.

\begin{figure}[ht]
	\centering
	\includegraphics[height=8.5cm]{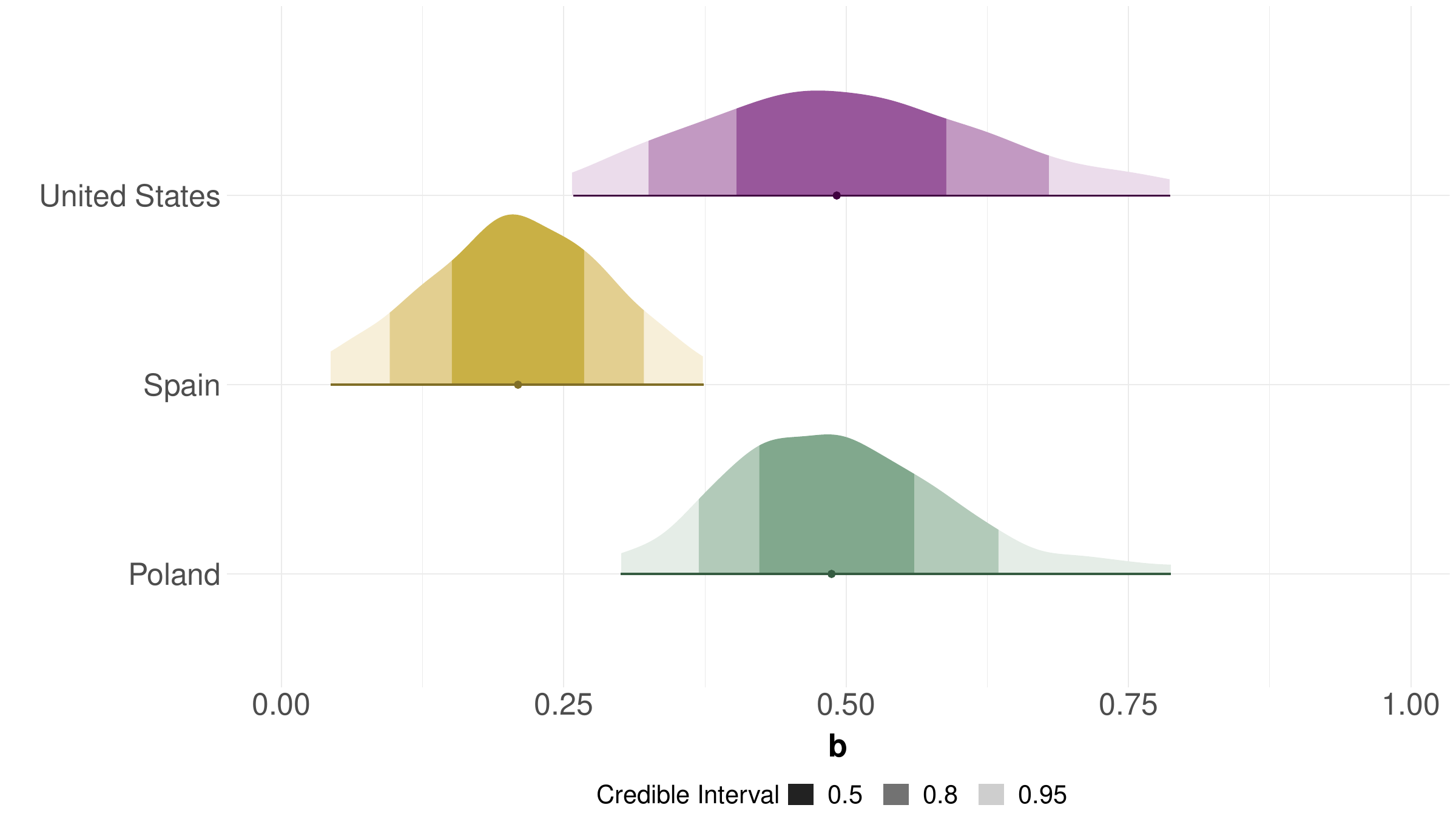}
	\caption{Comparison of the posterior vanishing parameter $b$ across countries. The point below the density represents the posterior mean.}
	\label{img: VanishingParam}
\end{figure}

Similar to the United States, the MA model assumes the years 2020 and 2021 to be jump years, as evident by the posterior estimates of $N_{2020}$ and $N_{2021}$ with mean values of one (cf. Figure~\ref{img: JumpOccurenceComp}). However, the effect of the pandemic was still present in 2022 as indicated by the estimate of the MA parameter~$b$ with a posterior mean of 0.5 (80\%-PI[0.363, 0.655]) shown in Figure~\ref{img: VanishingParam}. Moreover, the size of COVID shock was similar in severity to that of the other countries with a posterior mean for $\mu_Y$ of 1.19 (80\%-PI[0.3, 1.94]), while $\sigma_Y$ is estimated to be 1.16 (80\%-PI[0.128, 2.6]). The posterior distributions can be found in panel b) of Figure~\ref{img: JumpParamComp}. 

Especially the older population has been impacted the most by the COVID-19 pandemic as evident by the posterior estimates of $\beta_{x}^{(J)}$ depicted in Figure \ref{img: AgeMortPattern}. Here, we see lowest posterior mean for the young children aged 5-14 as well as a linear increase for higher ages with a maximum for the age group 75-84. Posterior estimates of all parameters including uncertainty quantification for both the AR and MA model can be found in Table~\ref{tb: Est_Pl_AR} and \ref{tb: Est_Pl_MA} in the Appendix. 

\subsection{Comparisons of pandemic effects across countries}

After having obtained estimates we can compare the results across countries. First, and most notably, the MA model consistently performs best, as evident by the lowest values of WAIC and of WAIC and LOO-CV (cf.~Table \ref{tb: WAIC_US}). We should note, that we also fit a MA-(2) model to all countries, however without improvement of the in-sample fit compared with the MA-(1) model. We therefore refrain from showing the results. The AR model, a choice where the initial shock affects subsequent periods longer, performs better than the Liu-Li model only for two the US and Spain. However, it can still be considered a good alternative. In terms of parameters estimates, the ones of the AR and MAR model are fairly similar. However, for sake of brevity we have decided not show these results in detail. 

Comparing posterior parameter estimates across various countries provides insights into the extent of their exposure to the COVID-19 pandemic. A compelling example is illustrated in Figure \ref{img: JumpOccurenceComp}, depicting the jump occurrences of all three countries. Notably, for Poland and the US, the initial pandemic shock has lasted for multiple periods, i.e. 2020 and 2021, before slowly vanishing, while that for Spain denotes a one-period event. The results are in line with that of other researchers who investigated excess mortality rates in Europe during the COVID pandemic and found high values in Spain for the year 2020 and Poland for both 2020 and 2021 \citep{bonnet2024}. Looking at the impact of COVID, we can compare the estimates of the intensity $Y_{2020}$ across all countries and see that the US has been affected the most followed by Poland and Spain. However, the impact of the second COVID wave in 2021 has been the highest in Poland. The posterior estimates of $Y_t$ can be found in panel a) of Figure~\ref{img: JumpParamComp}. 

Next to magnitude of the COVID pandemic, we can also compare the hyperparameters of the jump intensity by looking at the posterior estimates $\mu_Y$ and $\sigma_Y$ for each country (see panel b) of Figure~\ref{img: JumpParamComp}). The posterior mean of $\mu_Y$ is comparable across all countries, however the underlying uncertainty surrounding these estimates is distinct. For example, in Spain we see that the posterior distribution of $\mu_Y$ is the widest, which is due to there being a single jump only. With more jumps the posterior gets more weight by likelihood thereby decreasing uncertainty, as evident in Poland. Here, the posterior distribution is the sharpest because there are multiple jumps with similar intensity. The US is somewhere between these extremes. For $\sigma_Y$, we observe comparable results in that the estimate of Spain has the most variation and that of Poland the least.  

Furthermore, leveraging the estimated vanishing effect parameter~$b$, we can assess the pace of recovery from shocks across countries. As demonstrated in Figure \ref{img: VanishingParam}, a distinct recovery pattern emerges. In the case of Spain, the posterior estimate is moderate, whereas those for Poland and the United States can be considered high. The data indicates, that Spain has undergone a more rapid recovery, while the United States and Poland continues to face increased mortality due to COVID-19 even in later years. 

\begin{figure}
    \centering
    \includegraphics[height=8.5cm]{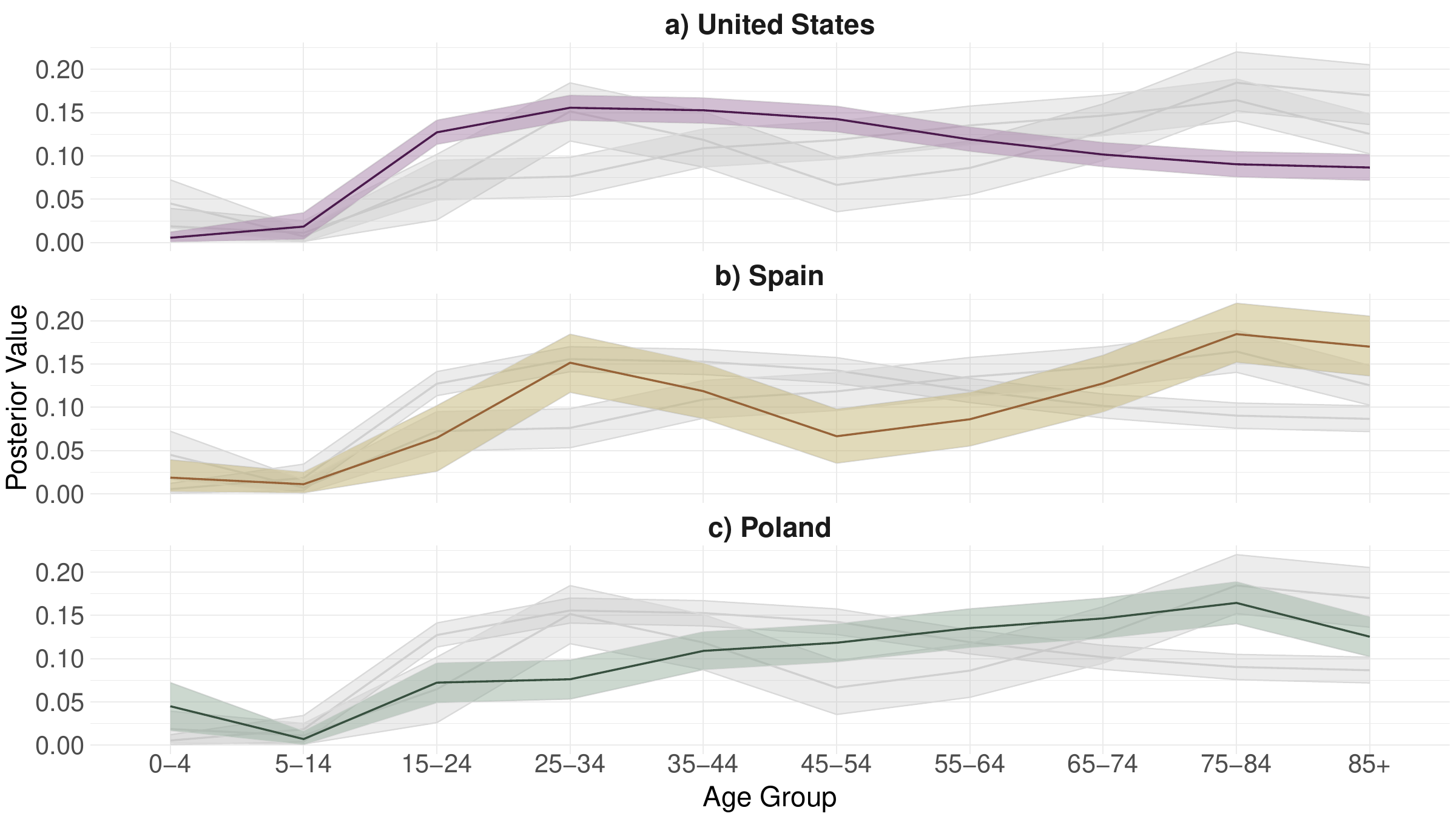}
    \caption{Posterior estimates of Jump effect $\beta_x^{(J)}$ for each country. Thick line denotes posterior mean while the shades denote 80\%- posterior intervals. Gray shaded area shows respective estimates of the other countries. }
    \label{img: AgeMortPattern}
\end{figure}

Lastly, the shock pattern exhibited by $\beta_x^{(J)}$ reveals distinct variations across different countries. To illustrate, in Poland, the shock predominantly affects the older population, whereas in the United States, the impact is distributed across a wide range of age groups, displaying a pronounced peak within middle age groups, as indicated by the posterior estimates. Spain occupies an intermediate position between the aforementioned patterns. This variance is visually depicted in Figure \ref{img: AgeMortPattern}, where the posterior mean shock pattern, along with its 80\%-PI, is depicted for each country across all age groups.

\subsection{Measuring the shock effect by age groups}

After having estimated all the parameters we were able to forecast future death rates. Moreover, we can calculate how much future death rates are affected by the addition of the age-specific jump effect $\beta_x^{(J)} \,J_{t}$. Using draws from the posterior predictive distribution we can calculate empirical quantiles for the shock component $\beta_x^{(J)} \,J_{t}$ to answer the question of how much of a percentage increase in death rates is likely to occur in the future due to a shock.

Hereby, consider that  the log death rate of our model consists of the log death rate of a LC model, i.e. a scenario without a jump, denoted $\ln \left(m_{x, t}^{LC}\right)$, plus the shock component $\beta_x^{(J)} \,J_{t}$ in case there is a jump:
\begin{equation}
    \ln \left(m_{x, t}\right)= \underbrace{\alpha_x \,+\, \beta_x\, \kappa_t\,+\, e_{x, t}}_{\ln \left(m_{x, t}^{LC}\right)} \,+\,  \underbrace{\beta_x^{(J)} \,J_{t}}_{\ln (c_{x.t})}. 
\end{equation}
Using basic rules of logarithms, we can calculate the percentage increase induced by the shock component to the death rate of a jump free scenario, with 
\begin{equation}
    m_{x, t} = m_{x, t}^{LC} \cdot c_{x,t}.
\end{equation}

\begin{figure}[ht]
    \centering
    \includegraphics[height=8.5cm]{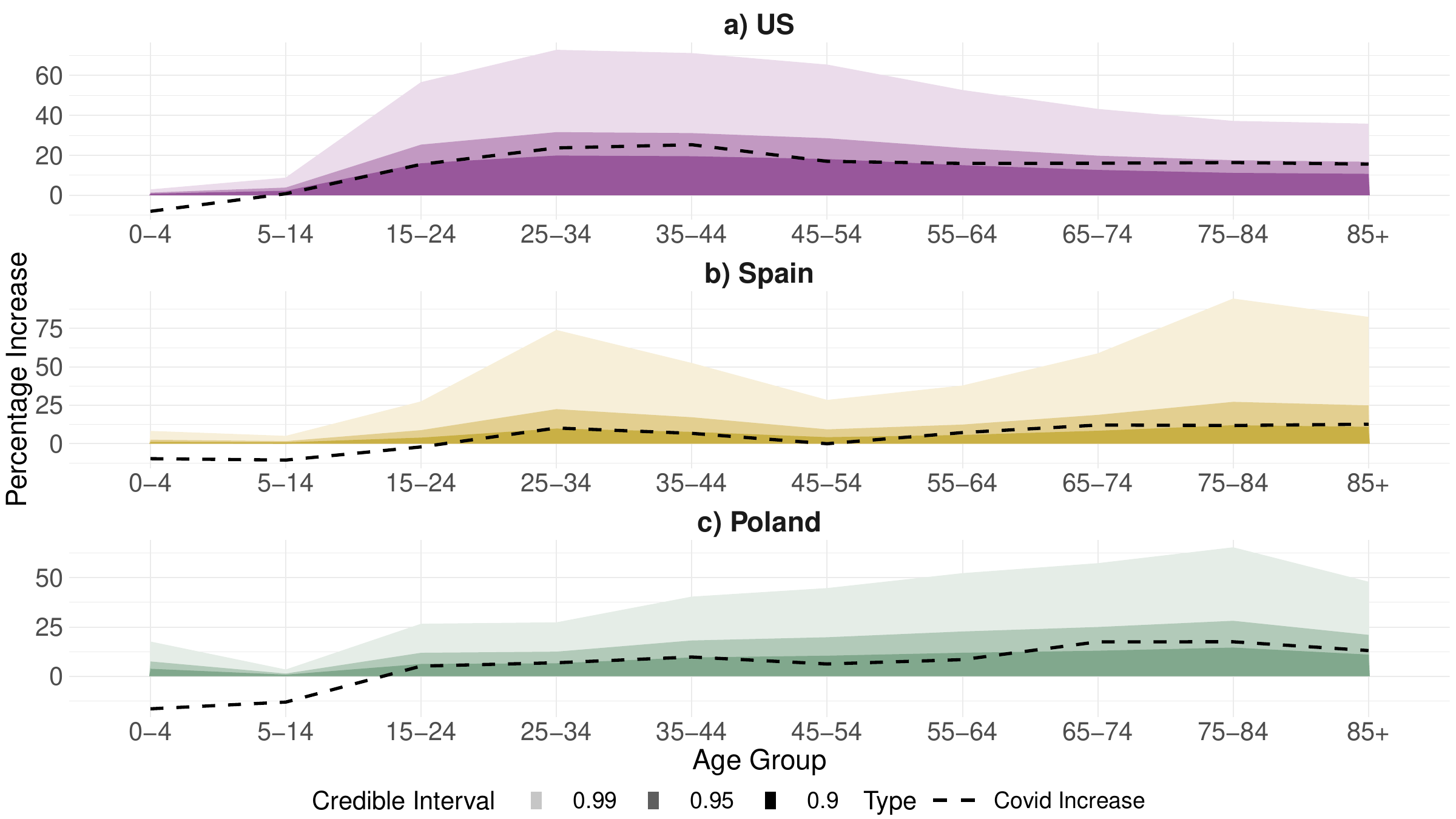}
    \caption{Observed percentage increase in death rates from the average of 2016 - 2019 to 2020 by countries (Covid Increase). The different shades denote the respective width of the prediction interval for $\bar{\tilde{c}}_{x} = \frac{1}{T}\sum^{T}_{t=1}\left[\exp{\beta_x J_t}-1\right]$.}
    \label{img: JumpAffection}
\end{figure}

In Figure \ref{img: JumpAffection}, we have plotted 90\%, 95\% and the 99\% credible interval of the predicted percentage increase $\tilde{c}_{x,t}=\exp(\beta_x J_t)-1$ by age group averaged over time. In addition we have added a dashed line showing the actual, observed increase to the death rates from due to the COVID-19 pandemic. Here, we have taken the average age specific death rates from 2016 to 2019 and calculated by how much this average was increased (or decreased) in 2020. We note that this increase is given in percentages, not percentage points. 

First, looking at Figure \ref{img: JumpAffection} we can see that the COVID-19 pandemic did not induce a parallel constant shift to the log death rates, as some authors \citep[e.g.][]{schnurch2022} proposed. Adding a constant to all age-specific log death rates would result in a constant percentage increase of the mortality rates, which is not observed. Second, using Figure \ref{img: JumpAffection}, we can provide an upper bound on the increase in mortality rates that our model predicts for a future time period. For example, our model states that for any future single year, the mortality rates for the age group 75-84 in Spain will not increase by more than 10\% with a probability of 95\%. However, we are aware that it is difficult to draw general conclusions for a future pandemic after having observed just one. The results should not be considered an attempt to forecast the severity of a future pandemic but rather as a tool to better visualise and capture the impact of the mortality shocks in the past. 


\section{Data analysis during the world wars: out-of-sample performance}\label{forecasting}

As mentioned in Section \ref{applications}, the recent occurrence of the pandemic does not allow for comparison of forecasting accuracy on out-of-sample data. Moreover, due to the rare nature of a mortality shock, i.e. there has not been another pandemic of this magnitude in the data set, we have to set the last jump indicator $N_T$ to zero to ensure model identifiability (see section~\ref{sec: ProofIdentification} in the Appendix), which makes forecasting during the pandemic difficult. There is however mortality data available for multiple countries in Europe during both World War I and World War II. Both wars, as well as the Spanish flu in 1918, can be considered as mortality shocks. We obtained mortality and population data for England and Wales from 1900 - 2000 from the HMD website and, for sake of consistency, divided the data into the same ten age groups as for the COVID data. We note that this is the same data used by \cite{liu2015age}. 


We compare the out-of-sample performance of our models on the war data using a training period (TRP) and a test period, distinguishing between two scenarios. Scenario one trains the parameters on the First World War and then predicts the future behaviour of mortality rates during the Second World War. Scenario two predicts future death rates after both world wars have occurred, that is during a time of low volatility. In the second scenario, we want to see if by specifically accounting for the two wars, i.e. the mortality shocks, we can reduce the uncertainty of the non-jump parameters and thus the uncertainty of our forecasts, without treating the extreme observations as outliers and removing them from the data, as is done, for example, by \cite{lee1992modeling}. This is of particular interest for forecasting mortality in the years following the COVID pandemic.

\subsection{Prediction of future death rates during times of war}

Following the methodology of Section \ref{ch: MortForecasts}, we can forecast future mortality rates recursively by application of Equation \eqref{eq: Projection}. We have trained the AR, the MA and the Liu-Li model on the England and Wales data for the years 1901 - 1943 and predict future death rates for a total of $H = 5$ years ahead. In this scenario we specifically choose to compare short term forecasts, as the models' forecasts will only differ significantly in the near future and we do not want to dilute the results. For longer ahead forecasts, all models will produce similar forecasts as shown in the second scenario below.  Moreover, in this scenario we do not have to assume that $N_T = 0$, since the vanishing parameters can be estimated during the period of the First World War, and we know of a time $\tilde{t}$ where there is no jump, e.g. 1930 (cf. section~\ref{sec: ProofIdentification}).

Examining the predictive performance of the Liu-Li model against the models that include some form of serial dependence in the jump component, we see that the latter models outperform the former by virtue of a lower score. In addition, the AR model, that is the model which carries the initial shock the longest in the future, performs best, an unsurprising result given the time of the last training observation. Out-of-sample results can be found in Table~\ref{tb: OOS-UK}.

\begin{table}[ht]
   \begin{center}
    \begin{minipage}{\textwidth}
			\caption{Out-of-sample fit comparison of the Liu-Li and our models on England and Wales data. Bold value denotes the best model for each column.}
			\label{tb: OOS-UK}
			\begin{tabularx}{\textwidth}{Xssss}
                \toprule  
                Score & AR & Liu-Li & MA & LC \\ 
                \midrule 
                \textbf{TRP = 1901-1943, H = 5} & & & & \\
                LogS & \textbf{-26.39} & -24.00 & -24.92 & - \\ 
                CRPS & \textbf{6.36} & 6.84 & 6.69 & - \\ 
                MSE (in percentages) & \textbf{8.55} & 10.51 & 9.44 & -\\ 
                MAE (in percentages) & 16.64 & 17.47 & \textbf{16.60} & - \\ 
                \vspace{0.1cm}
                \textbf{TRP = 1901-1980, H = 30}  & & & & \\
                LogS & \textbf{-68.00} & -67.48 & -67.78 & 128.46 \\ 
                CRPS & 32.59 & 32.74 & \textbf{32.55} & 68.25 \\ 
                MSE (in percentages) & \textbf{3.67} & 3.73 & 3.71 & 15.61 \\ 
                MAE (in percentages) & \textbf{14.49} & 14.63 & 14.61 & 31.72 \\ 
                  \bottomrule
			\end{tabularx}
  \raggedright
	\small{\textit{Notes:} TRP = training period. H = forecast horizon.}
   \end{minipage}
   \end{center}
\end{table}

\subsection{Prediction of future death rates during normal times}

In the second scenario, we trained all models on the England and Wales data for the years 1901 - 1980. In addition, we fit a standard LC model as well, where the parameters were estimated on the mortality improvement rates using the Route II approach. For all models, we created long term forecasts, i.e. future death rates for $H=30$ years ahead. As shown in Table~\ref{tb: OOS-UK}, the models that explicitly account for the past mortality shocks, that is the Liu-Li as well as the AR and MA model, produce superior forecasts in terms of lower scores. More precisely, the predictive accuracy of the jump models during times of low volatility is almost identical and the slight deviation can be attributed to sampling variation, as the results are all within less than 1\% of each other. Thus, in terms of forecasting, there is nothing to lose by choosing a more complex model such as the AR model over the Liu-Li model, even when forecasting in normal times. On the other hand, the predictive accuracy of the LC model is considerably lower, as the mortality shocks affect the parameter estimates of $\beta_x$ and $d$ substantially, while also increasing the variance of $(\xi_t)_t$ and $(\varepsilon_{x,t})_t$. Both effects significantly reduce the predictive power and highlight the need to explicitly account for past unusual mortality events, such as wars or pandemics, if accurate estimates of future mortality rates are to be obtained.  

\begin{figure}[h]
    \centering
    \includegraphics[height=7cm]{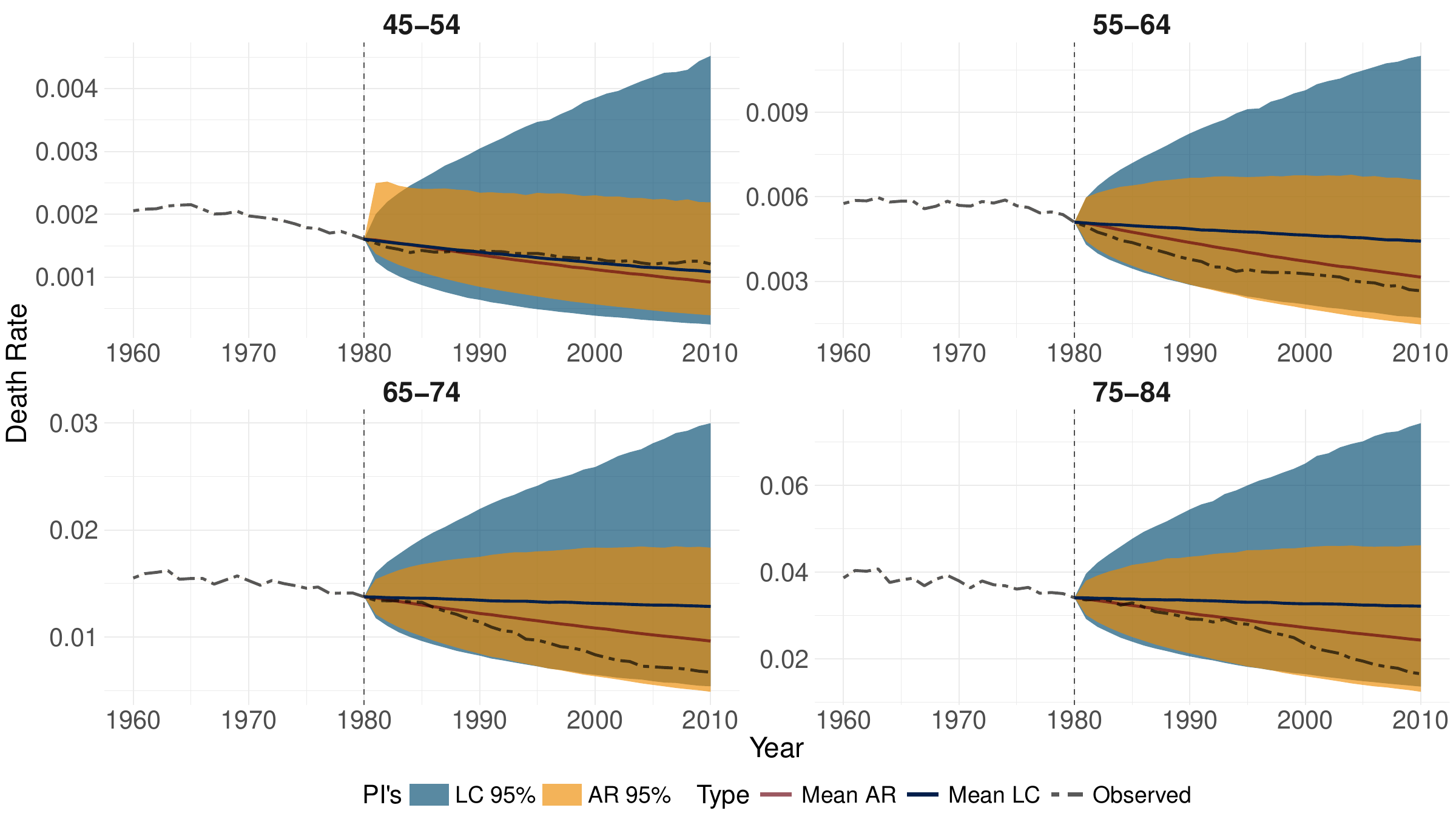}
    \caption{30-year ahead forecasts of death rates for different age groups of England and Wales including prediction intervals (PI) using both a LC and AR model. Dashed black line denotes actual observed values, while blue line denotes the mean forecast (Mean LC) of the LC model and red the mean forecast (Mean AR) of own model.}
    \label{img: ForecastsUK80}
\end{figure}

To visualise the above results, we compare the forecasts of the LC and our model in Figure~\ref{img: ForecastsUK80}.  Here, we notice the substantially larger prediction intervals of the LC model compared with those of the AR model.

Finally, we point out that prediction intervals of future death rates are used in an insurance context to determine the solvency capital for life insurance liabilities (see e.g. \cite{robben2023catastrophe}). Figure \ref{img: ForecastsUK80} illustrates that the predictions intervals from the Lee-Carter model are unrealistic and there is an important need to build a mortality model that adequately accounts for mortality shocks such as the models proposed in this paper.

\section{Extension: Multi-population mortality model with vanishing jumps}\label{sec:multipopulation}

Instead of estimating the model for single populations only, so called multi-population models provide more robust mortality forecasts for multiple populations that share similar socioeconomic characteristics. Its strength lies in the usage of more data by combination of different data  sources, i.e. countries, which increases stability.  \cite{li2005coherent} were the first to extend the classical LC model into a multi-population framework. A Bayesian implementation was introduced by \cite{antonio2015bayesian}. For an overview, we refer to \cite{enchev2017multi}. 

There are many possibilities of extending our proposed modelling framework into a multi-population setting. A detailed discussion of all of them would exceed the scope of this paper. Examples include a common factor model, where all countries share a common age and time effect \citep[e.g.][]{li2005coherent}. However, the addition of another age-time interaction term increases the amount of identifiability constraints needed. Alternatively, a two-step estimation approach may be considered \citep{antonio2015bayesian}. Another possible choice is the co-integrated LC model \citep[e.g.][]{li2011} where the time parameter $\kappa_t$ follows a multivariate random walk with drift.  

In the following, we propose a possible multi-population extension with a shared mortality jump occurrence parameter given by the following model 
\begin{equation} \label{eq: OwnMod_MultiPop}
    \begin{split}
        \ln(m_{x,t,c}) = \alpha_{x,c} + \beta_{x,c}\kappa_{t,c}+
        \beta_{x,c}^{(J)}J_{t,c}+\varepsilon_{x,t,c} 
    \end{split}
\end{equation}
with $c = \{1,2,\dots,C\}$ denoting the index for a given country, while  $\boldsymbol{\kappa}_{t}=\left(\kappa_{t,1},\dots,\kappa_{t,C}\right)^{\intercal}$ is modelled as a multivariate random walk with drift. We can use both the AR as well as the MA structure to model the serial dependence of the country specific jump component $J_{t,c}$. The model of \eqref{eq: OwnMod_MultiPop} using an AR structure is given by
\begin{equation*}
    J_{t,c} = a_{c}\,J_{t-1,c} + N_t\,Y_{t,c}.
\end{equation*}
Here, $N_t$ implies the existence of a global shock that affects all countries with local effect size $J_{t,c}$. Similar to single-population models proposed in Section \ref{modelling}, each country has its own vanishing parameter $a_c$ and jump severity $Y_{t,c}$. However, they share the same jump occurrence process $N_t$.  

Note that all identification conditions derived for single populations models directly carry over to the present setting. The only thing that needs to be checked is whether a model of the form \eqref{eq: OwnMod_MultiPop}, where $(\beta_{x,c})_x$ or $(\beta_{x,c}^{(J)})_x$ do not sum to 1, can be transformed into a model that satisfies these conditions without deviating from the joint value of the global parameter $N_t$. This is easily achieved by appropriate rescaling of the country-specific parameters, that is by using $\beta_{x,c}/(\sum_x\beta_{x,c})$, $\beta_{x,c}^{(J)}/(\sum_x\beta_{x,c}^{(J)})$, $\kappa_{t,c}\sum_x\beta_{x,c},$ and $Y_{t,c}\sum_x\beta_{x,c}^{(J)}$ instead of $\beta_{x,c}$, $\beta_{x,c}^{(J)}$, $\kappa_{t,c},$ and $Y_{t,c}$, respectively.

\subsection{In-sample comparison}

To compare single and multi-population models, we use the WAIC metric. 
When fitting single-population models separately, it is implicitly assumed that the parameters and likelihood of the single population models are independent of each other. Therefore, we want to compare the WAIC of the joint multi-population model that accounts for the dependence between countries with the sum of the WAIC scores of the single population models. Indeed, when assuming independence, the total WAIC can be decomposed into the sum of single-population WAICs as it is shown below.

Suppose some data $y$ with total sample size $n$ can be decomposed in two parts, where $I_1 = \{1,\dots, m\}$ and $I_2 = \{m+1,\dots, n\}$, with $i \in I_{1}$ and $ j \in I_{2} $ as well as corresponding parameters $\theta = (\theta_{1}, \theta_{2})^{'}$ that are independent of each other. Moreover let $ \theta_{1} $ denote the parameters of all $y_{i}$ and $ \theta_{2} $ those of all $ y_{j} $. The lpd can then be approximated as as 
\begin{align*}
	\widehat{{\text{lpd}}} &= \sum_{l = 1}^{n} \log \left( \frac{1}{S} \sum_{s = 1}^{S} 
	p(y_{l}|\theta^{(s)})
	\right) \\ 
	 &= \sum_{i \in I_1} \log \left( \frac{1}{S} \sum_{s = 1}^{S} 
	p(y_{i}|\theta_{1}^{(s)})\right) + 
	\sum_{j \in I_2} \log \left( \frac{1}{S} \sum_{s = 1}^{S} 
	p(y_{j}|\theta_{2}^{(s)})\right) \\ 
	 &= \widehat{{\text{lpd}}}_{1} + \widehat{{\text{lpd}}}_{2}. 
\end{align*}
In addition, the penalty term of the WAIC for the amount of parameters can be expressed as 
\begin{align*}
	\widehat{p}_{waic}  & = \sum_{l = 1}^{n} V_{s = 1}^{S} \log p(y_{l}|\theta^{(s)}) \\ 
 	& = \sum_{i \in I_1} V_{s = 1}^{S}  \log p(y_{i}|\theta^{(s)}_{1}) + 
 	\sum_{j \in I_{2}} V_{s = 1}^{S} \log p(y_{j}|\theta^{(s)}_{2}) \\
	& =  \widehat{p}_{waic}^{(1)} +\widehat{p}_{waic}^{(2)},  
\end{align*}
The above decomposition of both $\widehat{{\text{lpd}}}$ and $\widehat{p}_{waic}$ allows us to compare the WAIC of the joint multi-population model with the sum of WAIC's of the single-population model. 

The LOO-CV can be decomposed similarly granting us another measure to compare the in-sample fit of the single and multi-population models. Moreover, let $y_{I_1} = (y_{1},...,y_{m})^{\intercal}$  and $y_{I_2} = (y_{m+1},...,y_{n})^{\intercal}$ then we obtain: 
\begin{align*}
	\text{lpd}_{\text{loo}} = &\sum_{l = 1}^{N} \log p(y_{l}|y_{-l}) \\
	= &\sum_{l = 1}^{N} \log \int p(y_{l}|\theta) p(\theta|y_{-l}) d\theta \\
	= &\sum_{l = 1}^{N} \log \int \int p(y_{l}|\theta_{1},\theta_{2}) p(\theta_{1},\theta_{2}|y_{-l}) d\theta_{1} d\theta_{2} \\
	= &\sum_{i \in I_{1}} \log \int p(y_{i}|\theta_{1})p(\theta_{1}|y_{-i}) d\theta_{1}  \underbrace{\int p(\theta_{2}|y_{I_{2}}) d\theta_{2}}_{=1} + \\
     &\sum_{j \in I_{2}} \log \int p(y_{j}|\theta_{2})p(\theta_{2}|y_{-j}) d\theta_{2} \underbrace{\int p(\theta_{1}|y_{I_{1}}) d\theta_{1}}_{=1} \\
	   = & \text{ lpd}_{\text{loo}}^{(1)} + \text{lpd}_{\text{loo}}^{(2)}.
\end{align*}

\subsection{Results}

We applied the multi-population model to the original mortality data from 1991 to 2023 to estimate the parameters for the United States, Spain and Poland jointly. For the multi-population model, the posterior mean estimates of the jump occurrences $N_{2020}$ and $N_{2021}$ are estimated to be 1. That is, the model assumes that both of these years are shock years, similar to the single country estimates of the US and Poland. Looking at the WAIC for the AR model, the in-sample fit can be drastically improved by the multi-population approach (see Table~\ref{tb: WAIC_Joint}). The lpd of both the single as well as the multi-population models are very comparable, however the latter uses $3\cdot(T-3)-3$ fewer parameters (the $N_t$ parameters, excluding the constraints, for all three countries, minus three correlation parameters for $\Delta \boldsymbol{\kappa}_{t}$) resulting in a substantially lower WAIC. For the MA model we observe the same findings in that the log scores of the single and multi-population models are very comparable, however the latter approach uses substantially less parameters, resulting in a lower WAIC value. Looking at the LOO-CV metric we see a superior performance of the multi-population approach as well, with the MA extension obtaining lower scores than the AR approach.  

\begin{table}[ht]
   \begin{center}
    \begin{minipage}{\textwidth}
			\caption{In-sample fit comparison on COVID-19 data for the multi-population approach. Bold value denotes best of the column.}
			\label{tb: WAIC_Joint}
			\begin{tabularx}{\textwidth}{XYY}
				\toprule
				Model& Single Population & Multi Population   \\ 
				\midrule
				\textbf{WAIC} & &  \\ 
				AR & -3832.12 & \textbf{-3850.42}  \\ 
                MA & -3841.02 & \textbf{-3859.42}   \\ 
                                \vspace{0.1cm}
                \textbf{LOO-CV} & & \\ 
                AR & -3809.73 & \textbf{-3830.30} \\ 
                MA & -3819.76 & \textbf{-3839.80} \\ 
				\bottomrule
			\end{tabularx}
   \end{minipage}
   \end{center}
\end{table}


Another advantage of the multi-population approach lies in its property of a joint jump occurrence $N_t$. If we were to generate a single forecast using the multi-population approach a future pandemic occurs at the same time across all countries which is not necessarily the case for independent single-population models. 

\section{Conclusion}\label{conclusion}

In this paper we have introduced a new class of models that allows for more accurate modelling of mortality rates in the event of a shock. More precisely, we have extended the well-known LC model structure to allow for the inclusion of a serial dependent jump effect, where the effect of a shock is largest at the beginning and then gradually diminishes over time, offering large flexibility. Compared with the approach of \cite{liu2015age}, our models better captures the underlying pattern of the COVID-19 pandemic for Spain, US and Poland. Additionally, we have demonstrated that the jump auto-correlation structure is applicable to various shock scenarios, as evidence by the improved out-of-sample fit in the case of war-related data in England and Wales. Hereby, we could also show,
that we need to explicitly account for past shocks if we want to make accurate predictions of future mortality. Moreover, we have introduced a multi-population extension with a shared jump occurrence parameter. Finally, we have proven the identification of parameters for both the single-population and multi-population models. 

A valid point of criticism, nonetheless, relates to the considerable variability observed in the jump parameters, namely $\mu_Y$, $\sigma_Y$, and $a$, respectively $b$. Employing a Bayesian hierarchical modelling approach could potentially reduce this variability by pooling information across dimension. However, the general problem remains: No matter the approach, the model will have difficulty in estimating the parameters with only a few data points available. Though, with the availability of an increased number of time points and thus mortality shocks a substantial reduction in the standard deviation will be achieved. In general, our method's efficacy is most pronounced when a larger temporal scope is available, enabling more robust estimation of the parameters.  

On the other hand, an advantage of the Bayesian approach that we have not explored further is the use of expert opinion for the specific choice of hyperparameters in prior distributions. For example, \cite{zhou2021multi} incorporates expert opinion to simulate future mortality scenarios for events similar to the COVID outbreak. The same expert opinion could be used to set more informative priors, exploiting the the interplay between prior and likelihood to update the posterior. In data-rich scenarios the estimates are primarily influenced by the data, whereas in data-scarce settings, the influence of the prior, or expert opinion, becomes more pronounced. This approach may substantially reduce the posterior variability and can be seen as a middle ground between purely expert based and data-only estimation. 

Moreover, our models assumes that after a shock, the trend of mortality rates tends to return to their pre-shock trend driven by their constant drift. However, it is also possible that the shock has introduced either a new trend or baseline level. For example, after a severe pandemic, the population may be more alert to infectious diseases, leading to greater caution during the winter months. This change in behaviour could lead to lower levels of mortality after the pandemic. On the contrary, the impact might decrease but not disappear completely. Rather it could converge toward a general baseline level, which results in a permanent effect that can be compared to other causes of death such as the flu. Such scenarios which are briefly discussed in \cite{van2022estimating} are not considered by our approach but can constitute an interesting generalisation of our model.  

Furthermore, the shock of a pandemic's mortality can trigger a compensatory response. There is an argument that a pandemic accelerates the demise of those already in poor health. This type of scenario, while not observed, was especially discussed at the beginning of the COVID-19 pandemic \citep[e.g.][]{cairns2020impact}. Here it is believed that many of those who die during a pandemic would have died anyway in the near future, resulting in a slight decrease in the mortality rates among survivors. This contradicts our assumption of a pandemic effect that slowly vanishes over time, making our model unsuitable for this type of scenario.  


In practical applications, our model holds promise for actuarial contexts, particularly in determining solvency capital for mortality and longevity risk, which is imposed by supervisory authorities. The wider confidence intervals provided by our approach suggest that insurance companies may need to increase their capital reserves to safeguard against future pandemics and mortality shocks. This highlights the real-world significance and potential impact of our modeling framework on risk management in the insurance industry.

\section{Data availability}
For full replication of the results, we provide the code including data at our GitHub repository, available at \url{https://github.com/goesj/VanishingJumps}

\section{Acknowledgement and Funding}
Julius Goes would like to thank the Bamberg Graduate School of Social Sciences (BAGSS) for their support. Moreover, Julius Goes gratefully acknowledges financial support by the Oberfrankenstiftung (grant FP01054).

\bibliographystyle{agsm}

\newpage
\section{Appendix}

\subsection{Proof of Identification}\label{sec: ProofIdentification}

Identifiability of $\sigma_r^2$ is obvious.
The conditions~$\Delta J_{2}=0$, $\xi_2 =0$ and $\sum_x \beta_x=1$ yield $\sum_x \mathbb{E}\left(Z_{x,1}\right)=d$, i.e.~identifiability of the drift, which in turn implies that $\Delta \kappa_2=d$.
Starting from~\eqref{eq: LC2 Trans 1} with $\mathbf{B_x}=(\beta_x,\,\beta_x^{(J)})^{\intercal}$  and $\mathbf{K_t}=(\Delta \kappa_{t+1},\, \Delta J_{t+1})^\intercal$ we can verify identifiability of  $(\beta_x)_x,\; (\beta_x^{(J)})_x, \; (\Delta \kappa_t)_t$ and $(\Delta J_t)_t$ showing that the only possible choice for the matrix~$\mathbf A$ is the identity matrix if the
  constraints
\begin{equation}\label{eq: ConstraintsModel}
	\sum_{x=1}^{A} \beta_{x}=\sum_{x=1}^{A} \tilde \beta_{x} = 1, \quad  \sum_{x=1}^{A} \beta_{x}^{(J)}=\sum_{x=1}^{A} \tilde\beta_{x}^{(J)} = 1, \quad \Delta J_{1}=\Delta\tilde J_1 = 0, \quad \Delta \kappa_{1} =\Delta\tilde \kappa_1= d.
 \quad 
\end{equation}
are met.
Then, from 
$$
\begin{pmatrix}
d\\0
\end{pmatrix}\,=\, \mathbf{\tilde K_1}\,=\, 
\begin{pmatrix}
a_1&a_3\\a_2&a_4
\end{pmatrix}\, \mathbf K_1=\begin{pmatrix}
a_1&a_3\\a_2&a_4
\end{pmatrix}\begin{pmatrix}
d\\0
\end{pmatrix}
$$
we get 
$a_1=1$ and $a_2=0$ in view of our assumption $d \neq 0$. Similarly,
$$
\begin{pmatrix}
1\\1
\end{pmatrix}\,=\, \sum_x\mathbf{\tilde B}_x\,=\, 
\frac{1}{a_1\,a_4-a_2\, a_3}\begin{pmatrix}
a_4&-a_3\\-a_2&a_1
\end{pmatrix}\, \sum_x\mathbf{ B}_x=\frac{1}{a_1\,a_4-a_2\, a_3}\begin{pmatrix}
a_4&-a_3\\-a_2&a_1
\end{pmatrix}\,\begin{pmatrix}
1\\1
\end{pmatrix}
$$
yields
$${a_4-a_3}=a_1-a_2=a_1\,a_4-a_2\,a_3.
$$
Hence, we end up with $a_3=0$ and $a_4=1$. In other words, we achieved identifiability of the parameters of the mortality improvements.

In other words, we achieved identifiability of the parameters of the mortality improvements. If we assume additionally that $J_1=J_2=0$, we can identify all of $(J_t)_t$ iteratively using $(\Delta J_t)_t$ as well as $\Delta J_2 = 0$. \\

Subsequently, we will discuss identifiablity of the parameters of the two proposed time series models for the jump processes separately. 

\subsubsection{Identification of Parameters in AR(1) Model}

To identify all other parameters in the AR(1) model, we have to assume that $N_T=0$ and that $(Y_t)_t$ is a sequence of positive parameters. This allows us to identify the times of jump occurencies $(N_t)_t$, the autoregressive parameter $a$ and jump intensity $(Y_t)_t$. Note that if there is no jump, then the parameter $a$ cannot be identified and can be neglected in the model. Otherwise we can proceed iteratively: Let $t^*$ denote the time of the first jump event, then we have $N_t=0, J_t=0$ for all $t<t^*$ as well as $N_{t^*}=1$ and $Y_{t^*}=J_{t^*}$. 
Noting that a jump event at time $t^*$ implies that $J_t>0$ for all $t\geq t^*$ if $a>0$, we can deduce that $a=0$ for $J_t=0$ for some $t>t^*$. For $a \neq 0$, we have $J_T=a J_{T-1}$ as $N_T=0$ by assumption. Hence, we can identify $a$ and subsequently $(N_t)_t$ and the $Y_t$'s belonging to non-vanishing $N_t$'s. 
 
As a final remark, let us mention that instead of assuming $N_T=0$ we could also assume that we know a time point $\tilde t>t^*$, where no jump occurs. Then the whole argument above can be adapted using $\tilde t$ instead of $T$. Moreover, instead of assuming positivity of $(Y_t)_t$ one can alternatively assume that $J_{T-1}\neq 0$.

\subsubsection{Identification of Parameters in MA(1) Model}

Identification of parameters in the MA(1) model follows a similar structure. Again we assume that $N_T=0$ and that $(Y_t)_t$ is a sequence of positive parameters. Let $t^*$ denote the first year of a jump which can be identified by observation of $J_{t^*}>0$. Then $N_{t^*} = 1$, $Y_{t^*} = J_{t^*}$ and $N_t = J_t = 0$ for all $t < t^*$. Note that  $(N_t Y_t)_{t\geq t^*}$ forms a non-homogeneous first order difference  equation with time-varying coefficients which can be solved recursively:
\begin{equation}
N_{t^*+h}Y_{t^*+h}= (-b)^h J_{t^*}+\sum_{k=1}^h (-b)^{h-k}J_{t^*+k-1},\quad h=1,\dots, T-t^*.
\end{equation}
Hence, if we are able to identify $b$, all remaining parameters are identified. To this end, recall that $N_T=0$ by assumption which gives
\begin{equation}\label{eq.id-ma}
0= (-b)^h J_{t^*}+\sum_{k=1}^{T-t^*} (-b)^{h-k}J_{t^*+k-1}.
\end{equation}
This means, that a unique root of the polynomial (in $b$) on the rhs in the interval [0,1) assures identifiability of $b$. 

For illustrative purposes, we can alternatively identify $b$ using the following step-by-step procedure with $t^*$ being the first year of a jump: First, if $J_{t^* + 1} < J_{t^*}$, then we can identify two sub-cases. If $J_{t^* + 2} = 0$ it follows that $ N_{t^*+1} = 0$ and we can estimate $b$, by noting that $J_{t^* +1} = b N_{t^*}Y_{t^*}$. However, if $J_{t^* + 2} \neq 0$ then either $N_{t^*+1} = 1$ or $N_{t^*+2} = 1$. The latter can be checked by seeing if $J_{t^* + 4} = 0$ allowing us to estimate $b$ using $J_{t^*+3}$. In case of the former, identification of the parameters is achieved similarly as described next.

Second, if $J_{t^* + 1} > J_{t^*}$, then $N_{t^*+1} = 1$, since $b \in [0,1)$. We then check if $N_{t^*+2} = 0$ by observation of $J_{t^*+4}$. If this is indeed the case $J_{t^*+1}$ can be rewritten to obtain $N_{t^*+1}Y_{t^*+1} = J_{t^*+1} - bN_{t^*}Y_{t^*}$ which can be substituted into $J_{t^*+2}$. Then we solve for $b$ and obtain the following quadratic expression 
\begin{equation*}
    b_{1/2} = \frac{-J_{t^*+1} \pm \sqrt{
    (J_{t^*+1})^2 - 4\cdot(-N_{t^*}Y_{t^*})\cdot(-J_{t^*+2})
    }}{2\cdot(-N_{t^*}Y_{t^*})}.    
\end{equation*}
We obtain a unique solution for $b$ by checking which of $b_{1}$ or $b_{2} \in [0,1)$ or by observation of a second shock at some later time period onward.

Again as in the AR-case, instead of assuming $N_T=0$, we could also assume that we know a time point $\tilde t>t^*$, where no jump occurs. One simply has to substitute the upper index $T-t^*$ of the sum in \eqref{eq.id-ma} by $\tilde t-t^*$. In particular, this leads to a reduction of the degree of the polynomial on the rhs which in turn simplifies the problem of finding its roots.


\subsection{Deriving Dirichlet distributions from Gamma distributions}\label{sec: DirichAsGamma}

Let $X_i$ be a random variable from the Gamma distribution with $X_i \sim  \operatorname{Gamma}(\alpha_i,1)$, where $i = 1,\dots, k$. Further, let 
\begin{equation}\label{eq: TransGamma}
    Y_i = \frac{X_i}{X_1+X_2+\dots+X_k}.
\end{equation}
Then, the joint density of $Y_1,\dots,Y_{k-1}$ is
	\begin{equation}\label{eq: jointPDFNormGamma}
		f(y_1,\dots,y_{k-1}) = \frac{\alpha_1+\dots + \alpha_k}{\Gamma(\alpha_1) \cdots \Gamma(\alpha_k)}y_{1}^{\alpha-1} \cdots y_{k-1}^{\alpha_{k-1}-1}\left(1-y_{1}-\dots-y_{k-1}\right)^{\alpha_{k}-1},
	\end{equation}
where $y_{i}>0, i = 1,\dots,k-1, y_{1}+ \dots + y_{k-1}<1$. The above joint pdf of 
$Y_1,\dots,Y_{k-1}$ happens to be the pdf of a Dirichlet distribution with parameters $\alpha_1,\dots,\alpha_k$ for the random vector $(Y_{1},\dots,Y_{k})$, with $y_k=1-y_{1}-\dots-y_{k-1}$. For proof see e.g. \cite{ng2011}.  

Hence, instead of sampling $\mathbf{\beta}$ from a Dirichlet distribution, we can also sample Gamma distributed random variables $b_{1},\dots,b_{A}$ and apply the transformation of \eqref{eq: TransGamma} by setting 
\begin{equation*}
    \beta_x = \frac{b_x}{b_{1}+\dots+b_{A}}, \text{ for all } x \in \{1,\dots,A-1\}
\end{equation*}
and $\beta_{A}=(1-\beta_{1} - \cdots \beta_{A-1})$. This allows us to use a wider variety of samplers, like the multivariate slice sampler of \cite{tibbits2014}, improving convergence.  

\newpage

\subsection{Tables of parameter estimates}
\subsubsection{Parameter Estimates of AR models for COVID Data}

\begin{table}[ht]
\begin{center}
\begin{minipage}{\linewidth}
\centering
\caption{Posterior estimates for the US data using the AR model} \label{tb: Est_US_AR}
\begin{tabularx}{\textwidth}{YXXXXXXXXX}
  \toprule  & Mean & MAP & Sd & 10\%-PI & 90\%-PI & split-$\hat{R}$ & Bulk-ESS & Tail-ESS \\ 
  \midrule $\beta_{1}$ & 0.11 & 0.10 & 0.02 & 0.08 & 0.13 & 1.00 & 1996 & 1852 \\ 
  $\beta_{2}$ & 0.16 & 0.16 & 0.02 & 0.13 & 0.19 & 1.00 & 2035 & 1736 \\ 
  $\beta_{3}$ & 0.14 & 0.14 & 0.02 & 0.11 & 0.17 & 1.00 & 1982 & 1884 \\ 
  $\beta_{4}$ & 0.19 & 0.19 & 0.02 & 0.16 & 0.22 & 1.00 & 1970 & 2050 \\ 
  $\beta_{5}$ & 0.16 & 0.16 & 0.02 & 0.13 & 0.19 & 1.00 & 1595 & 1977 \\ 
  $\beta_{6}$ & 0.08 & 0.08 & 0.02 & 0.05 & 0.11 & 1.00 & 1847 & 1887 \\ 
  $\beta_{7}$ & 0.07 & 0.07 & 0.02 & 0.05 & 0.10 & 1.00 & 1975 & 1941 \\ 
  $\beta_{8}$ & 0.05 & 0.05 & 0.02 & 0.02 & 0.08 & 1.00 & 2075 & 2006 \\ 
  $\beta_{9}$ & 0.02 & 0.01 & 0.02 & 0.00 & 0.04 & 1.00 & 2036 & 2048 \\ 
  $\beta_{10}$ & 0.01 & 0.00 & 0.01 & 0.00 & 0.03 & 1.00 & 2036 & 1759 \\ 
  $\beta^{(J)}_{1}$ & 0.01 & 0.00 & 0.00 & 0.00 & 0.01 & 1.00 & 1926 & 1792 \\ 
  $\beta^{(J)}_{2}$ & 0.02 & 0.01 & 0.01 & 0.00 & 0.03 & 1.00 & 1825 & 1782 \\ 
  $\beta^{(J)}_{3}$ & 0.13 & 0.13 & 0.01 & 0.11 & 0.14 & 1.00 & 1994 & 1791 \\ 
  $\beta^{(J)}_{4}$ & 0.16 & 0.16 & 0.01 & 0.14 & 0.17 & 1.00 & 1832 & 1911 \\ 
  $\beta^{(J)}_{5}$ & 0.15 & 0.15 & 0.01 & 0.14 & 0.17 & 1.00 & 1616 & 2006 \\ 
  $\beta^{(J)}_{6}$ & 0.14 & 0.14 & 0.01 & 0.13 & 0.16 & 1.00 & 1842 & 1824 \\ 
  $\beta^{(J)}_{7}$ & 0.12 & 0.12 & 0.01 & 0.11 & 0.13 & 1.00 & 1896 & 1967 \\ 
  $\beta^{(J)}_{8}$ & 0.10 & 0.10 & 0.01 & 0.09 & 0.12 & 1.00 & 1920 & 2006 \\ 
  $\beta^{(J)}_{9}$ & 0.09 & 0.09 & 0.01 & 0.08 & 0.10 & 1.00 & 1916 & 1852 \\ 
  $\beta^{(J)}_{10}$ & 0.09 & 0.08 & 0.01 & 0.07 & 0.10 & 1.00 & 1827 & 1549 \\ 
  $d$ & -0.09 & -0.09 & 0.04 & -0.14 & -0.05 & 1.00 & 2166 & 2004 \\ 
  $\sigma_\xi$ & 0.20 & 0.19 & 0.03 & 0.16 & 0.25 & 1.00 & 1565 & 1857 \\ 
  $\sigma_{r}$ & 0.02 & 0.02 & 0.00 & 0.02 & 0.02 & 1.00 & 2028 & 1925 \\ 
  $p$ & 0.06 & 0.05 & 0.03 & 0.02 & 0.11 & 1.00 & 1819 & 1695 \\ 
  $a$ & 0.39 & 0.41 & 0.07 & 0.30 & 0.48 & 1.00 & 1359 & 1119 \\ 
  $\mu_{Y}$ & 1.31 & 1.39 & 0.68 & 0.42 & 2.07 & 1.00 & 1632 & 1266 \\ 
  $\sigma_{Y}$ & 1.07 & 0.27 & 0.88 & 0.19 & 2.28 & 1.00 & 218 & 259 \\ 
   \bottomrule  
   \end{tabularx}
   \end{minipage}
   \end{center}
\end{table}

\newpage

\begin{table}[ht]
\begin{center}
\begin{minipage}{\linewidth}
\centering
\caption{Posterior estimates for Spain using the AR model} \label{tb: Est_Sp_AR}
\begin{tabularx}{\textwidth}{YXXXXXXXXX}
  \toprule  & Mean & MAP & Sd & 10\%-PI & 90\%-PI & split-$\hat{R}$ & Bulk-ESS & Tail-ESS \\ 
  \midrule $\beta_{1}$ & 0.12 & 0.12 & 0.02 & 0.09 & 0.14 & 1.01 & 2001 & 1964 \\ 
  $\beta_{2}$ & 0.16 & 0.16 & 0.02 & 0.14 & 0.19 & 1.00 & 1933 & 1899 \\ 
  $\beta_{3}$ & 0.18 & 0.18 & 0.02 & 0.16 & 0.21 & 1.00 & 2071 & 1964 \\ 
  $\beta_{4}$ & 0.15 & 0.16 & 0.02 & 0.13 & 0.18 & 1.00 & 1818 & 1634 \\ 
  $\beta_{5}$ & 0.11 & 0.11 & 0.02 & 0.09 & 0.13 & 1.00 & 1983 & 1838 \\ 
  $\beta_{6}$ & 0.04 & 0.04 & 0.02 & 0.02 & 0.06 & 1.00 & 1998 & 1938 \\ 
  $\beta_{7}$ & 0.04 & 0.05 & 0.02 & 0.02 & 0.07 & 1.00 & 1901 & 1649 \\ 
  $\beta_{8}$ & 0.06 & 0.06 & 0.02 & 0.04 & 0.08 & 1.00 & 2008 & 1940 \\ 
  $\beta_{9}$ & 0.07 & 0.07 & 0.02 & 0.05 & 0.09 & 1.00 & 1865 & 1849 \\ 
  $\beta_{10}$ & 0.06 & 0.06 & 0.02 & 0.04 & 0.08 & 1.00 & 1676 & 1889 \\ 
  $\beta^{(J)}_{1}$ & 0.02 & 0.00 & 0.02 & 0.00 & 0.04 & 1.00 & 1941 & 1964 \\ 
  $\beta^{(J)}_{2}$ & 0.01 & 0.00 & 0.01 & 0.00 & 0.02 & 1.00 & 2077 & 1940 \\ 
  $\beta^{(J)}_{3}$ & 0.06 & 0.07 & 0.03 & 0.03 & 0.10 & 1.00 & 1888 & 1938 \\ 
  $\beta^{(J)}_{4}$ & 0.15 & 0.15 & 0.03 & 0.12 & 0.19 & 1.00 & 1904 & 2096 \\ 
  $\beta^{(J)}_{5}$ & 0.12 & 0.12 & 0.02 & 0.09 & 0.15 & 1.00 & 1992 & 1858 \\ 
  $\beta^{(J)}_{6}$ & 0.07 & 0.06 & 0.02 & 0.04 & 0.10 & 1.00 & 2208 & 1957 \\ 
  $\beta^{(J)}_{7}$ & 0.09 & 0.09 & 0.02 & 0.06 & 0.12 & 1.00 & 1961 & 1908 \\ 
  $\beta^{(J)}_{8}$ & 0.13 & 0.13 & 0.03 & 0.10 & 0.16 & 1.00 & 2087 & 1966 \\ 
  $\beta^{(J)}_{9}$ & 0.18 & 0.18 & 0.03 & 0.15 & 0.22 & 1.00 & 1518 & 1608 \\ 
  $\beta^{(J)}_{10}$ & 0.17 & 0.17 & 0.03 & 0.13 & 0.20 & 1.00 & 2107 & 1887 \\ 
  $d$ & -0.28 & -0.28 & 0.05 & -0.34 & -0.21 & 1.00 & 2027 & 1893 \\ 
  $\sigma_\xi$ & 0.32 & 0.31 & 0.05 & 0.26 & 0.39 & 1.00 & 1591 & 1724 \\ 
  $\sigma_{r}$ & 0.03 & 0.03 & 0.00 & 0.03 & 0.04 & 1.00 & 1979 & 1941 \\ 
  $p$ & 0.05 & 0.02 & 0.03 & 0.01 & 0.09 & 1.00 & 1601 & 1887 \\ 
  $a$ & 0.29 & 0.27 & 0.11 & 0.15 & 0.43 & 1.00 & 1770 & 1690 \\ 
  $\mu_{Y}$ & 1.19 & 1.15 & 1.01 & 0.21 & 2.52 & 1.00 & 802 & 967 \\ 
  $\sigma_{Y}$ & 1.32 & 0.61 & 1.00 & 0.26 & 2.71 & 1.00 & 224 & 223 \\ 
   \bottomrule 
   \end{tabularx}
   \end{minipage}
   \end{center}
\end{table}

\newpage

\begin{table}[ht]
\begin{center}
\begin{minipage}{\linewidth}
\centering
\caption{Posterior estimates for Poland using the AR model} \label{tb: Est_Pl_AR}
\begin{tabularx}{\textwidth}{YXXXXXXXXX}
  \toprule  & Mean & MAP & Sd & 10\%-PI & 90\%-PI & split-$\hat{R}$ & Bulk-ESS & Tail-ESS \\ 
   \midrule $\beta_{1}$ & 0.19 & 0.19 & 0.02 & 0.17 & 0.22 & 1.00 & 1684 & 1726 \\ 
  $\beta_{2}$ & 0.19 & 0.19 & 0.02 & 0.16 & 0.22 & 1.00 & 2000 & 1933 \\ 
  $\beta_{3}$ & 0.11 & 0.11 & 0.02 & 0.08 & 0.13 & 1.00 & 1887 & 1926 \\ 
  $\beta_{4}$ & 0.09 & 0.10 & 0.02 & 0.07 & 0.12 & 1.00 & 1883 & 1825 \\ 
  $\beta_{5}$ & 0.09 & 0.09 & 0.02 & 0.06 & 0.11 & 1.00 & 1910 & 1699 \\ 
  $\beta_{6}$ & 0.08 & 0.08 & 0.02 & 0.05 & 0.10 & 1.00 & 1740 & 1904 \\ 
  $\beta_{7}$ & 0.08 & 0.07 & 0.02 & 0.05 & 0.10 & 1.00 & 2102 & 1826 \\ 
  $\beta_{8}$ & 0.05 & 0.05 & 0.02 & 0.03 & 0.08 & 1.00 & 1794 & 1851 \\ 
  $\beta_{9}$ & 0.06 & 0.06 & 0.02 & 0.04 & 0.09 & 1.00 & 2098 & 2029 \\ 
  $\beta_{10}$ & 0.06 & 0.06 & 0.02 & 0.03 & 0.08 & 1.00 & 1809 & 1829 \\ 
  $\beta^{(J)}_{1}$ & 0.04 & 0.05 & 0.02 & 0.02 & 0.07 & 1.00 & 2038 & 1740 \\ 
  $\beta^{(J)}_{2}$ & 0.01 & 0.00 & 0.01 & 0.00 & 0.02 & 1.00 & 1959 & 1924 \\ 
  $\beta^{(J)}_{3}$ & 0.07 & 0.07 & 0.02 & 0.05 & 0.09 & 1.00 & 1994 & 1774 \\ 
  $\beta^{(J)}_{4}$ & 0.08 & 0.08 & 0.02 & 0.05 & 0.10 & 1.00 & 1885 & 1958 \\ 
  $\beta^{(J)}_{5}$ & 0.11 & 0.11 & 0.02 & 0.09 & 0.13 & 1.00 & 1987 & 1857 \\ 
  $\beta^{(J)}_{6}$ & 0.12 & 0.12 & 0.02 & 0.10 & 0.14 & 1.00 & 1823 & 1902 \\ 
  $\beta^{(J)}_{7}$ & 0.14 & 0.13 & 0.02 & 0.11 & 0.16 & 1.00 & 2045 & 1751 \\ 
  $\beta^{(J)}_{8}$ & 0.15 & 0.15 & 0.02 & 0.12 & 0.17 & 1.00 & 2016 & 1966 \\ 
  $\beta^{(J)}_{9}$ & 0.16 & 0.16 & 0.02 & 0.14 & 0.19 & 1.00 & 2118 & 1915 \\ 
  $\beta^{(J)}_{10}$ & 0.13 & 0.13 & 0.02 & 0.10 & 0.15 & 1.00 & 1917 & 1888 \\ 
  $d$ & -0.26 & -0.26 & 0.05 & -0.32 & -0.20 & 1.00 & 1996 & 1761 \\ 
  $\sigma_\xi$ & 0.25 & 0.24 & 0.05 & 0.20 & 0.31 & 1.00 & 1829 & 1962 \\ 
  $\sigma_{r}$ & 0.03 & 0.03 & 0.00 & 0.03 & 0.04 & 1.00 & 1976 & 1888 \\ 
  $p$ & 0.06 & 0.05 & 0.03 & 0.02 & 0.11 & 1.00 & 1790 & 1898 \\ 
  $a$ & 0.33 & 0.35 & 0.09 & 0.23 & 0.43 & 1.00 & 1216 & 830 \\ 
  $\mu_{Y}$ & 1.29 & 1.37 & 0.64 & 0.45 & 1.93 & 1.01 & 1397 & 915 \\ 
  $\sigma_{Y}$ & 0.95 & 0.16 & 0.86 & 0.12 & 2.21 & 1.03 & 147 & 238 \\ 
   \bottomrule  
   \end{tabularx}
   \end{minipage}
   \end{center}
\end{table}

\newpage

\subsubsection{Parameter Estimates of MA models for COVID Data}
\begin{table}[ht]
\begin{center}
\begin{minipage}{\linewidth}
\centering
\caption{Posterior estimates for the US data using the MA model} \label{tb: Est_US_MA}
\begin{tabularx}{\textwidth}{YXXXXXXXXX}
  \toprule  & Mean & MAP & Sd & 10\%-PI & 90\%-PI & split-$\hat{R}$ & Bulk-ESS & Tail-ESS \\ 
  \midrule $\beta_{1}$ & 0.10 & 0.10 & 0.02 & 0.07 & 0.13 & 1.00 & 2019 & 1872 \\ 
  $\beta_{2}$ & 0.15 & 0.15 & 0.02 & 0.12 & 0.18 & 1.00 & 1866 & 1964 \\ 
  $\beta_{3}$ & 0.14 & 0.14 & 0.02 & 0.11 & 0.16 & 1.00 & 1906 & 1749 \\ 
  $\beta_{4}$ & 0.19 & 0.20 & 0.02 & 0.17 & 0.22 & 1.00 & 1995 & 1850 \\ 
  $\beta_{5}$ & 0.17 & 0.16 & 0.02 & 0.14 & 0.19 & 1.00 & 2008 & 1748 \\ 
  $\beta_{6}$ & 0.08 & 0.08 & 0.02 & 0.06 & 0.11 & 1.00 & 2048 & 1743 \\ 
  $\beta_{7}$ & 0.08 & 0.08 & 0.02 & 0.05 & 0.10 & 1.00 & 2141 & 1921 \\ 
  $\beta_{8}$ & 0.05 & 0.05 & 0.02 & 0.03 & 0.08 & 1.00 & 1891 & 1885 \\ 
  $\beta_{9}$ & 0.02 & 0.02 & 0.02 & 0.00 & 0.05 & 1.00 & 2153 & 1961 \\ 
  $\beta_{10}$ & 0.01 & 0.00 & 0.01 & 0.00 & 0.03 & 1.00 & 1912 & 1984 \\ 
  $\beta^{(J)}_{1}$ & 0.01 & 0.00 & 0.00 & 0.00 & 0.01 & 1.00 & 1824 & 1679 \\ 
  $\beta^{(J)}_{2}$ & 0.02 & 0.00 & 0.01 & 0.00 & 0.03 & 1.00 & 1916 & 2005 \\ 
  $\beta^{(J)}_{3}$ & 0.13 & 0.13 & 0.01 & 0.11 & 0.14 & 1.00 & 1721 & 1925 \\ 
  $\beta^{(J)}_{4}$ & 0.16 & 0.16 & 0.01 & 0.14 & 0.17 & 1.00 & 2073 & 2046 \\ 
  $\beta^{(J)}_{5}$ & 0.15 & 0.15 & 0.01 & 0.14 & 0.17 & 1.00 & 2199 & 1780 \\ 
  $\beta^{(J)}_{6}$ & 0.14 & 0.14 & 0.01 & 0.13 & 0.16 & 1.00 & 2056 & 1924 \\ 
  $\beta^{(J)}_{7}$ & 0.12 & 0.12 & 0.01 & 0.11 & 0.13 & 1.00 & 2074 & 1846 \\ 
  $\beta^{(J)}_{8}$ & 0.10 & 0.10 & 0.01 & 0.09 & 0.12 & 1.00 & 2057 & 1751 \\ 
  $\beta^{(J)}_{9}$ & 0.09 & 0.09 & 0.01 & 0.08 & 0.11 & 1.00 & 1661 & 1886 \\ 
  $\beta^{(J)}_{10}$ & 0.09 & 0.09 & 0.01 & 0.07 & 0.10 & 1.00 & 1706 & 1962 \\ 
  $d$ & -0.09 & -0.09 & 0.04 & -0.14 & -0.05 & 1.00 & 2002 & 2010 \\ 
  $\sigma_\xi$ & 0.22 & 0.20 & 0.04 & 0.17 & 0.26 & 1.00 & 1586 & 1920 \\ 
  $\sigma_{r}$ & 0.02 & 0.02 & 0.00 & 0.02 & 0.02 & 1.00 & 1889 & 1725 \\ 
  $p$ & 0.09 & 0.05 & 0.05 & 0.03 & 0.15 & 1.00 & 2025 & 1789 \\ 
  $b$ & 0.50 & 0.48 & 0.14 & 0.33 & 0.68 & 1.01 & 844 & 900 \\ 
  $\mu_{Y}$ & 1.15 & 1.26 & 0.78 & 0.24 & 2.02 & 1.00 & 1617 & 1887 \\ 
  $\sigma_{Y}$ & 1.27 & 0.71 & 0.88 & 0.37 & 2.47 & 1.00 & 525 & 493 \\ 
  \bottomrule 
   \end{tabularx}
   \end{minipage}
   \end{center}
\end{table}

\newpage

\begin{table}[ht]
\begin{center}
\begin{minipage}{\linewidth}
\centering
\caption{Posterior estimates for Spain using the MA model} \label{tb: Est_Sp_MA}
\begin{tabularx}{\textwidth}{YXXXXXXXXX}
  \toprule  & Mean & MAP & Sd & 10\%-PI & 90\%-PI & split-$\hat{R}$ & Bulk-ESS & Tail-ESS \\ 
  \midrule $\beta_{1}$ & 0.11 & 0.11 & 0.02 & 0.09 & 0.13 & 1.00 & 2015 & 1884 \\ 
  $\beta_{2}$ & 0.16 & 0.16 & 0.02 & 0.14 & 0.18 & 1.00 & 2084 & 1879 \\ 
  $\beta_{3}$ & 0.18 & 0.19 & 0.02 & 0.16 & 0.20 & 1.00 & 1882 & 1852 \\ 
  $\beta_{4}$ & 0.15 & 0.15 & 0.02 & 0.13 & 0.18 & 1.00 & 2090 & 1766 \\ 
  $\beta_{5}$ & 0.11 & 0.11 & 0.02 & 0.09 & 0.13 & 1.00 & 2137 & 2018 \\ 
  $\beta_{6}$ & 0.04 & 0.04 & 0.02 & 0.02 & 0.06 & 1.00 & 1981 & 1923 \\ 
  $\beta_{7}$ & 0.05 & 0.04 & 0.02 & 0.03 & 0.07 & 1.00 & 1906 & 1924 \\ 
  $\beta_{8}$ & 0.06 & 0.06 & 0.02 & 0.04 & 0.08 & 1.00 & 1855 & 1923 \\ 
  $\beta_{9}$ & 0.07 & 0.07 & 0.02 & 0.05 & 0.09 & 1.00 & 2037 & 1881 \\ 
  $\beta_{10}$ & 0.06 & 0.06 & 0.02 & 0.04 & 0.08 & 1.00 & 1688 & 1886 \\ 
  $\beta^{(J)}_{1}$ & 0.02 & 0.00 & 0.02 & 0.00 & 0.04 & 1.00 & 2151 & 1989 \\ 
  $\beta^{(J)}_{2}$ & 0.01 & 0.00 & 0.01 & 0.00 & 0.02 & 1.00 & 2047 & 1899 \\ 
  $\beta^{(J)}_{3}$ & 0.06 & 0.06 & 0.03 & 0.03 & 0.10 & 1.00 & 1796 & 1833 \\ 
  $\beta^{(J)}_{4}$ & 0.16 & 0.16 & 0.03 & 0.12 & 0.19 & 1.00 & 2008 & 1882 \\ 
  $\beta^{(J)}_{5}$ & 0.12 & 0.12 & 0.03 & 0.09 & 0.15 & 1.00 & 1879 & 1924 \\ 
  $\beta^{(J)}_{6}$ & 0.07 & 0.07 & 0.02 & 0.03 & 0.10 & 1.00 & 1919 & 1947 \\ 
  $\beta^{(J)}_{7}$ & 0.09 & 0.08 & 0.03 & 0.05 & 0.12 & 1.00 & 1973 & 1942 \\ 
  $\beta^{(J)}_{8}$ & 0.13 & 0.12 & 0.03 & 0.10 & 0.16 & 1.00 & 2013 & 1967 \\ 
  $\beta^{(J)}_{9}$ & 0.18 & 0.18 & 0.03 & 0.15 & 0.22 & 1.00 & 1993 & 1938 \\ 
  $\beta^{(J)}_{10}$ & 0.17 & 0.16 & 0.03 & 0.13 & 0.20 & 1.00 & 1835 & 1858 \\ 
  $d$ & -0.28 & -0.28 & 0.05 & -0.34 & -0.21 & 1.00 & 2038 & 1655 \\ 
  $\sigma_\xi$ & 0.33 & 0.32 & 0.05 & 0.27 & 0.40 & 1.00 & 1910 & 1938 \\ 
  $\sigma_{r}$ & 0.03 & 0.03 & 0.00 & 0.03 & 0.04 & 1.00 & 1923 & 1814 \\ 
  $p$ & 0.05 & 0.03 & 0.03 & 0.01 & 0.09 & 1.00 & 1488 & 1414 \\ 
  $b$ & 0.21 & 0.20 & 0.09 & 0.10 & 0.32 & 1.00 & 1881 & 1855 \\ 
  $\mu_{Y}$ & 1.19 & 0.19 & 1.06 & 0.17 & 2.52 & 1.00 & 406 & 663 \\ 
  $\sigma_{Y}$ & 1.34 & 0.62 & 1.02 & 0.33 & 2.84 & 1.01 & 174 & 217 \\ 
  \bottomrule 
   \end{tabularx}
   \end{minipage}
   \end{center}
\end{table}

\newpage

\begin{table}[ht]
\begin{center}
\begin{minipage}{\linewidth}
\centering
\caption{Posterior estimates for Poland using the MA model} \label{tb: Est_Pl_MA}
\begin{tabularx}{\textwidth}{YXXXXXXXXX}
  \toprule  & Mean & MAP & Sd & 10\%-PI & 90\%-PI & split-$\hat{R}$ & Bulk-ESS & Tail-ESS \\ 
  \midrule $\beta_{1}$ & 0.20 & 0.20 & 0.02 & 0.17 & 0.23 & 1.00 & 1964 & 1964 \\ 
  $\beta_{2}$ & 0.20 & 0.20 & 0.02 & 0.17 & 0.23 & 1.00 & 1882 & 1883 \\ 
  $\beta_{3}$ & 0.11 & 0.11 & 0.02 & 0.08 & 0.13 & 1.00 & 2062 & 2006 \\ 
  $\beta_{4}$ & 0.09 & 0.09 & 0.02 & 0.07 & 0.12 & 1.00 & 2028 & 2005 \\ 
  $\beta_{5}$ & 0.09 & 0.09 & 0.02 & 0.06 & 0.11 & 1.00 & 2165 & 2093 \\ 
  $\beta_{6}$ & 0.08 & 0.08 & 0.02 & 0.05 & 0.10 & 1.00 & 2019 & 1844 \\ 
  $\beta_{7}$ & 0.07 & 0.07 & 0.02 & 0.05 & 0.10 & 1.00 & 2119 & 2050 \\ 
  $\beta_{8}$ & 0.05 & 0.05 & 0.02 & 0.02 & 0.07 & 1.00 & 1859 & 1885 \\ 
  $\beta_{9}$ & 0.06 & 0.06 & 0.02 & 0.04 & 0.09 & 1.00 & 2057 & 2007 \\ 
  $\beta_{10}$ & 0.05 & 0.06 & 0.02 & 0.03 & 0.08 & 1.00 & 2002 & 1981 \\ 
  $\beta^{(J)}_{1}$ & 0.05 & 0.05 & 0.02 & 0.02 & 0.07 & 1.00 & 1736 & 1661 \\ 
  $\beta^{(J)}_{2}$ & 0.01 & 0.00 & 0.01 & 0.00 & 0.02 & 1.00 & 1799 & 1709 \\ 
  $\beta^{(J)}_{3}$ & 0.07 & 0.07 & 0.02 & 0.05 & 0.10 & 1.00 & 2171 & 1900 \\ 
  $\beta^{(J)}_{4}$ & 0.08 & 0.07 & 0.02 & 0.05 & 0.10 & 1.00 & 1890 & 1886 \\ 
  $\beta^{(J)}_{5}$ & 0.11 & 0.11 & 0.02 & 0.09 & 0.13 & 1.00 & 1919 & 1819 \\ 
  $\beta^{(J)}_{6}$ & 0.12 & 0.12 & 0.02 & 0.10 & 0.14 & 1.00 & 1864 & 1927 \\ 
  $\beta^{(J)}_{7}$ & 0.14 & 0.14 & 0.02 & 0.11 & 0.16 & 1.00 & 1741 & 1851 \\ 
  $\beta^{(J)}_{8}$ & 0.14 & 0.15 & 0.02 & 0.12 & 0.17 & 1.00 & 1755 & 1988 \\ 
  $\beta^{(J)}_{9}$ & 0.16 & 0.17 & 0.02 & 0.14 & 0.19 & 1.00 & 2109 & 2141 \\ 
  $\beta^{(J)}_{10}$ & 0.13 & 0.13 & 0.02 & 0.10 & 0.15 & 1.00 & 1943 & 1887 \\ 
  $d$ & -0.26 & -0.25 & 0.05 & -0.31 & -0.20 & 1.00 & 2123 & 1956 \\ 
  $\sigma_\xi$ & 0.25 & 0.24 & 0.04 & 0.20 & 0.31 & 1.00 & 1918 & 2005 \\ 
  $\sigma_{r}$ & 0.03 & 0.03 & 0.00 & 0.03 & 0.04 & 1.00 & 2094 & 2007 \\ 
  $p$ & 0.06 & 0.04 & 0.03 & 0.02 & 0.11 & 1.00 & 1628 & 1985 \\ 
  $b$ & 0.50 & 0.48 & 0.12 & 0.37 & 0.64 & 1.00 & 1943 & 1423 \\ 
  $\mu_{Y}$ & 1.22 & 1.28 & 0.57 & 0.50 & 1.71 & 1.01 & 1064 & 399 \\ 
  $\sigma_{Y}$ & 0.72 & 0.09 & 0.78 & 0.06 & 1.80 & 1.02 & 154 & 233 \\ 
   \bottomrule
   \end{tabularx}
   \end{minipage}
   \end{center}
\end{table}

\newpage



\subsection{Prior parameterisation for COVID data}\label{S95}

\begin{table}[h]
\begin{center}
\begin{minipage}{\linewidth}
    \caption{Prior parameterisation for all countries using COVID data (US, Spain and Poland).}\label{tb: PriorUS}
	\begin{tabularx}{\linewidth}{XYYY}
			\toprule
			Parameter& Prior Distribution  & Hyperprior 1  &  Hyperprior 2  \\
			\midrule & & & \\
            \vspace{-0.6cm}
			Age Parameter & & &  \\
			$\left(\beta_{1},\dots,\beta_{A}\right)$ & Dirichlet($ 1,\dots,1 $) & - & -  \\
			$\left(\beta_{1}^{(J)},\dots,\beta_{A}^{(J)}\right)$& Dirichlet($ 1,\dots,1 $)  & - & -  \\
			\vspace{0.05cm}
			Time Parameter & &  & \\ 
			$ \Delta \kappa_{t} $  & $\Delta \kappa_{t} \stackrel{iid}{\sim}\mathcal{N}(d,\sigma^{2}_{\xi})$ & $ d \sim \mathcal{N}(0,5^{2}) $ & $ \sigma_{\xi} \sim \mathcal{N}_{+}(0,2^{2}) $   \\
			$ N_{t} $& $ N_{t} \stackrel{iid}{\sim}\text{Bern}(p) $   & $ p \sim \text{Beta}(1,20) $  & -  \\
			$ Y_t$ &  $Y_t\stackrel{iid}{\sim}\mathcal{N}_{+}(\mu_{Y},\sigma^{2}_{Y})$ &  $\mu_{Y} \sim \mathcal{N}_{+}(0,4^{2})$  &   $\sigma_{Y} \sim \mathcal{N}_{+}(0,2^{2})  $ \\ 
			$ a $ & $ a \sim  \mathcal{N}_+(0,0.4^2) $ & - & -  \\
			\vspace{0.05cm}
			Other Parameters & & &  \\
			$ \sigma_{r} $ & $ \sigma_{r} \sim \mathcal{N}_{+}(0,2^{2})$ & - & - \\
			\bottomrule
		\end{tabularx}
  \end{minipage}
\medskip
\begin{minipage}{\linewidth}
	\caption{Prior parameterisation for the England and Wales Data}\label{tb: PriorUK}
	\begin{tabularx}{\linewidth}{XYYY}
		\toprule
		Parameter& Prior Distribution  & Hyperprior 1  &  Hyperprior 2  \\
		\midrule & & & \\
		\vspace{-0.6cm}
		Age Parameter & & &  \\
		$\left(\beta_{1},\dots,\beta_{A}\right)$ & Dirichlet($ 1,\dots,1 $)  & - & -  \\
		$\left(\beta_{1}^{(J)},\dots,\beta_{A}^{(J)}\right)$& Dirichlet($ 0.5,0.5,0.5,5,5,5,5,0.5,0.5,0.5 $) & - & -  \\
		\vspace{0.05cm}
		Time Parameter & &  & \\ 
		$ \Delta \kappa_{t} $  & $\Delta \kappa_{t} \stackrel{iid}{\sim}\mathcal{N}(d,\sigma^{2}_{\xi})$ & $ d \sim \mathcal{N}(0,5^{2}) $ & $ \sigma_{\xi} \sim \mathcal{N}_{+}(0,2^{2}) $   \\
		$ N_{t} $& $ N_{t} \stackrel{iid}{\sim}\text{Bern}(p) $   & $ p \sim \text{Beta}(1,20) $  & -  \\
		$ Y_t$ &  $Y_t\stackrel{iid}{\sim}\mathcal{N}(\mu_{Y},\sigma^{2}_{Y})$ &  $\mu_{Y} \sim \mathcal{N}_{+}(0,5^{2})$  &   $\sigma_{Y} \sim \mathcal{N}_{+}(0,5^{2})  $ \\ 
		$ a $ & $ a \sim  \text{Beta}(1,5) $ & - & -  \\
		\vspace{0.05cm}
		Other Parameters & & &  \\
		$ \sigma_{\varepsilon} $ & $ \sigma_{\varepsilon} \sim \mathcal{N}_{+}(0,2^{2})$ & - & - \\
		\bottomrule
	\end{tabularx}
\end{minipage}
  \end{center}
\end{table}

\pagebreak

\newpage
 
\subsection{Overview on used samplers for own model}\label{S96}

\begin{table}[h!]
\begin{center}
\begin{minipage}{\linewidth}
    \caption{Overview on selected samplers for own model}\label{tb: Overview Samplers}
		\begin{tabularx}{\linewidth}{XY}
			\toprule
			Sampler& Parameter  \\
			\midrule
			AF Slice Sampler &  \\
			& 	$\left(Y_{3},\dots,Y_{T-1},\mu_Y, \sigma_Y\right)$     \\
			Binary Sampler  &    \\ 
			& $ N_{t} \quad  \forall t \in \{1,\dots, T-1\} $    \\
			Gibbs  &   \\ 
			& $d$  \\
			& $p$  \\
			Random Walk Metropolis  &    \\
			& $\sigma_{r}$  \\
			& $\sigma_{\xi}$  \\
			Slice Sampler  &    \\ 
            & $a$ \\
            HMC & \\ 
            & 
            $\left(\Delta \kappa_{2},\dots,\Delta \kappa_{T},
            \beta_{1},\dots,\beta_{A} , \beta_{1}^{(J)},\dots,\beta_{A}^{(J)} \right)$ \\ 
         \bottomrule
	\end{tabularx}
 \end{minipage}
 \end{center}
\end{table}

\pagebreak

\end{document}